\documentclass[12pt,preprint]{aastex}

\usepackage{graphicx}
\usepackage{epstopdf}
\usepackage{cite}
\usepackage{hyperref}
 \usepackage{snapshot}
\graphicspath{oldplots,Lucy_Plots,blc_plots,bb_dev}
\begin{document}

\title{Bolometric and UV Light Curves of Core-Collapse Supernovae} 

\shorttitle{Swift UV observations of Core Collapse Supernovae}
\shortauthors{Pritchard et al.}

\author{T.~A.~Pritchard$^1$, P.~W.~A.~Roming$^{2,1}$,~Peter~J~Brown$^3$,~Amanda~J.~Bayless$^2$, ~Lucille~H.~Frey$^{4,5}$} \email{}
\affil{$^1$ Department of Astronomy \& Astrophysics, Penn State University, 525 Davey Lab, University Park, PA 16802, USA \\
	$^2$ Southwest Research Institute, Department of Space Science, 6220 Culebra Rd, San Antonio, TX 78238, USA \\
	$^3$ Department of Physics and Astronomy, George P. and Cynthia Woods Mitchell Institute for Fundamental Physics \& Astronomy, Texas A. \& M. University, 4242 TAMU, College Station, TX 77843, USA \\
	$^4$ Los Alamos National Laboratory, Los Alamos, NM 87545, USA \\
	$^5$ Department of Computer Science, University of New Mexico, Albuquerque, NM 87131, USA\\}
\begin{abstract}
The {\em Swift} UV-Optical Telescope (UVOT) has been observing Core-Collapse Supernovae (CCSNe) of all subtypes in the UV and optical since 2005.  We present here 50 CCSNe observed with the {\em Swift} UVOT, analyzing their UV properties and behavior.  Where we have multiple UV detections in all three UV filters  ($\lambda_c = 1928 - 2600$ \AA ), we generate early time bolometric light curves, analyze the properties of these light curves, the UV contribution to them, and derive empirical corrections for the UV-flux contribution to optical-IR based bolometric light curves.  

\end{abstract}

\section{Introduction}\label{Intro}
For decades, nearby Type Ia supernovae (SNe) have been extensively studied from the optical to near-IR (NIR) wavelength range \citep{Filippenko97,Krisciunas04,Wood-Vasey07}. UV observations, on the other hand, are historically much more limited and mostly include a handful of bright events from IUE and HST \citep{Panagia03}. More recently, the sample of nearby Type Ia SNe studied in the UV has dramatically increased \citep{Foley08,Brown09,Brown10,Cooke11,Maguire12b}.

In contrast to Type Ia SNe, core collapse supernovae (CCSNe) have not received the same level of attention. With the emergence of dedicated SN follow-up programs and telescopes, such as the Katzman Automatic Imaging Telescope \citep{Filippenko01}, Carnegie Supernova Project \citep{Hamuy06}, Fred Lawrence Whipple Observatory \citep{Matheson08}, PeterÕs Automated Infrared Imaging Telescope \citep{Bloom06}, Caltech Core Collapse Program\citep{Gal-Yam07b}, Palomar Transient Factory \citep{Rau09,Law09}, and the efforts of the Center for Astrophysics Supernova (SN) Group, nearby CCSNe are now being more frequently monitored in both the optical and NIR wavelength ranges with ground-based telescopes.

Despite this surge of interest, UV studies of nearby CCSNe have lagged behind redder wavelengths even though the UV is a promising probe of these interesting objects. This lack of UV observations is primarily due to the fact that UV studies blue-ward of the U-band are limited by the availability of space-based UV telescopes. Previous to 2005, 17 CCSNe were observed in the UV, primarily by the IUE and HST instruments \citep[see][for a complete census of pre-2005 UV observations]{Brown09}. Efforts to interpret these observations have underscored the utility of UV observations to better understand CCSNe events.

The physics governing a CCSN light curve is the time-scale and wavelength dependence of the diffusion of photons as radiation is transported towards the ÒsurfaceÓ to escape \citep{Hoeflich96}. The resultant light curves for stripped envelope CCSNe (Type Ib/c \& IIb) are principally due to the radioactive decay of $^{56}$Ni$\rightarrow ^{56}$Co$\rightarrow ^{56}$Fe \citep{Tominaga05}. Observationally, we may break CCSNe down into several subtypes depending upon observed light curve and spectral characteristics \citep{Filippenko05} and which are thought to have progenitor main sequence stars primarily differentiated by mass \citep{Smartt09}.  Type II SNe, that is SNe with hydrogen in their spectra, are broken down into IIn SNe which exhibit narrow hydrogen emission lines, IIP which have a long lived ($\sim$ 100 day) optical plateau, IIL which have a linear light curve decline after peak brightness, and IIb which show hydrogen soon after explosion and then rapidly evolve with weakening H lines and the development of He lines.  Type Ib/c SNe are relatively similar in that their spectra show no hydrogen but may or may not show helium lines for Ib/c respectively.  From a physical standpoint these differences are all thought to be related to the mass of the progenitor and the amount of hydrogen envelope remaining upon explosion.  For hydrogen-rich envelope CCSNe (i.e. Type IIP/L/n) the primary energy source is shock deposited followed by hydrogen recombination in the ejecta. Unlike stripped CCSNe, variations in energy input due to $^{56}$Ni mass and its associated radioactive decay in Type IIP SNe do not significantly affect the light curve shape, but serve instead to modify the plateau lifetime by a few weeks \citep{Kasen09}. Emergent spectra are dominated by continuum emission with a complex collection of absorption and emission lines bearing evidence of various elements in the optically thin surface region. Recently \citet{Dessart10}, using non-LTE time-dependent radiative-transfer modeling of a CCSNe, that the evolving UV spectrum is primarily driven by line blanketing and metallicity dependencies. The timing and depth of the iron-peak absorption is thus considered a probe of the amount of these elements near the surface.

Since 2005, the NASA {\em Swift} mission \citep{Gehrels04} has dramatically improved the number of CCSNe observed in the UV  as well as Type Ia SNe ($\sim80$). The {\em Swift} satellite has a 30cm Ultraviolet/Optical Telescope \citep[UVOT;][]{Roming05} capable of observing in three UV filters (central wavelengths; uvw2: $\lambda_c = 1928$ \AA; uvm2: $\lambda_c = 2246$ \AA; uvw1:$ \lambda_c = 2600$ \AA), three optical filters (u, b, v), and a UV and optical grism \citep{Poole08}.  Figure 2 of \citet{Poole08} provides more information on the filter response curves. The primary mission of the {\em Swift} satellite is to detect and monitor gamma ray bursts (GRBs); all SNe science performed with UVOT is secondary to that mission.  However, just as SNe are discovered after explosion via ÒblindÓ searches rather than observations of a known location, the isotropically distributed GRBs must also be detected via blind searches. This isotropic distribution means that {\em Swift} can point in the direction of, and observe, any particular SN without affecting the chances of a GRB discovery or the GRB response time. In this sense, UVOT is an ideal UV monitoring instrument with its rapid response to targets of opportunity (ToO) and the ease of submitting observation requests for them. While UVOT may not have the sensitivity or resolution of the Hubble Space Telescope (HST), these attributes allow UVOT to respond to SNe days faster than HST and obtain  more numerous individual observations making it the workhorse instrument of UV CCSNe studies.

In this paper, we present UV observations of CCSNe as observed by the NASA {\em Swift} satellite from launch through early 2012.  In Section \ref{obs} we discuss the SNe observations and data reduction pipeline.  In Section \ref{UVLC} we examine the properties of the sample light curves and their associated colors.  We calculate observed absolute magnitudes, color evolution and UV decay rate/light curve shape, as well as examine differences in these values based upon SN subtype.  In Section \ref{BLC} we use a particularly well observed subset of this sample to calculate bolometric light curves for these SNe at early times where the UV flux is a sizable fraction of the total luminosity.  We examine these based upon SN subtype, and from these light curves we calculate UV-bolometric corrections based on optical colors for use as an empirical correction to ground based optical-IR CCSNe pseudo-bolometric light curves.  

\section{Observations}\label{obs}
{\em Swift} observations of CCSNe are triggered as ToO observations, typically after a SNe candidate is found via other surveys and reported in the Central Bureau for Electronic Telegrams (CBET), International Astronomical Union Circulars (IAUC) or Astronomer's Telegram (ATEL).  Observations of CCSNe are commonly proposed by a number of different science working groups; however, in order to leverage UVOT's UV capabilities most observed SNe have the following characteristics:  (1) Low line of sight galactic reddening (E(B$-$V$) \lesssim 0.03$), (2) $ \ge 10\arcsec$ separation from the host galactic core to minimize coincidence losses due to a bright background, (3) Nearby (z $ \lesssim 0.02$), and  (4) SNe thought to be discovered `early' such that UV detections are likely.  This typically means either a recent pre-explosion upper limit, an observed spectrum with a strong blue continuum, or a best match photometrically or spectrally with a young CCSNe.   These are of course guidelines, not search criteria, and have developed over the course of the mission and often been ignored in the case of uniquely interesting events.  This suggests that our sample as presented here is biased, but as the largest sample available we use it to draw some broad conclusions about the UV behavior of these objects. 

Once a target of opportunity has been triggered, {\em Swift} usually commences observations typically using six color filters.  A typical observational cadence will vary over the campaign with observations often starting with a short separation of $\sim 1-2$ days as we examine the early emission and identify UV variability.  The cadence then typically lengthens out to $\sim 1$ week as we begin to lose UV detections and a greater integration time is needed.  A follow up $6-10$ ks observation for galaxy host-light template subtraction is often observed $\sim 0.5-1$ year later if no prior observations of the host galaxy have occurred.  A summary of SN observed by {\em Swift} and included in this sample may be seen in Table \ref{sne_table}.  A typical exposure time ranges from 2 ks at early times when the object is bright to $ 4-6$ ks as the SNe fade, and the number of observations vary from $\sim 6-50$.  SNe observations without any clear UV detections have been excluded from this paper.  

Images have been obtained from the NASA HEASARC {\em Swift} Archive.  All Images have been processed from the raw image and event files using the recent observations and calibrations, and all photometry measurements performed in this paper have been performed using NASA Heasoft v6.12.  Aspect corrections were performed manually when the automated processing pipeline failed, and images that were unable to be corrected have been excluded.  {\em Swift} has an approximately 96 minute orbit, of which a maximum of only $\sim30$ mins can be spent observing a single target due to scheduling constraints such as other observations, telescope pointing constraints (due to the sun, earth, and solar panel illumination), temperature and momentum constraints.   As such, an individual SNe's observations are often spread over multiple orbits, in which case each observation (i.e. segment) was co-added over all orbits to generate a single image.  Only limited co-adding was done outside of this to keep each observation within a short and well defined timespan of 1 day.  On occasions when a SN was bright enough to warrant UVOT grism observations, a short UVOT single-filter photometric observation occurred as part of the spectral observation.  If a detection occurred in that short snapshot it is reported individually, and thus there may be multiple observations in the same filter on a different observation ID overlapping in time (if the grism orbits were interspersed with the photometric orbits in the observation schedule).  Exposure (EXP) and Large Scale Structure (LSS) maps were generated for each processed sky file  and co-added along with the source sky image to be used in the photometry pipeline outlined below.

The {\em Swift} UVOT is a photon-counting device and as such there are several differences when compared with a typical optical CCD instrument that must be taken into account when performing photometry upon SNe.  The primary concern is coincidence loss of photons due to multiple photons arriving during the detector's readout time (which is similar to pileup as seen in X-ray CCDs).  Coincidence loss is non-linear above a certain count rate and while the correction for this  has been well calibrated for field objects, especially bright point sources, sources on a galaxy background require some special consideration.  We follow the basic photometry recipe from \citet{Brown09} designed to account for these particular challenges that SNe present, with some modifications as discussed below to account for an updated instrument calibration and pipeline.  We continue to perform aperture photometry using a $3\arcsec$ aperture.  This is smaller than the $5 \arcsec$ aperture recommended by \citet{Poole08} for use on isolated objects, but due to the fact that this is on the same scale as the UVOT point spread function it has been empirically found to lower the contamination to the bright background of the host galaxy \citep{Li06}.  We account for sensitivity variations across the detector via the incorporation of LSS and EXP maps, as well as the mission time dependent sensitivity loss,  into the {\tt uvotsoursce} pipeline as discussed in the updated UVOT calibrations in \citet{Breeveld10}.   

A $5\arcsec$ aperture is used to determine the coincidence loss rate so that we remain consistent with the instrumental calibrations, and we add a 3\% uncertainty in quadrature with the Poisson errors in order to estimate the uncertainty due to small scale structure \citep{Brown09,Poole08}.  This is a conservative estimate as it is unchanged from previous papers before the advent of the \citet{Breeveld10} calibrations.    Where we have a pre-explosion image or a suitably late observation ($\gtrsim $ 6 months  - 1 year), we subtract the galaxy count rate from the SN + galaxy observations.  The ability to obtain these observations is constrained by {\em Swift}'s heavy subscription rate, and therefore of our 49 objects this has been performed for all filters for 28 SNe, in the UV filters only for 3 SNe, and not at all for 18 SNe.  See Table \ref{sne_table} for an individual SN's status.  The effect of a missing template image varies  - when the SNe are much brighter than the host galaxy the effect is minimal; however, for faint SNe missing these observations results in the possibility of spurious detections, systematically brighter observations, and a shallower slope than would otherwise be observed.  Using our sample observations that have been template subtracted, we compare photometry before and after this process in order to to examine the effect that this has upon our data.  This may be seen in Table \ref{templ_table}, where we show the mean, standard deviation and maximum difference in magnitudes that the template subtraction process corrects for due to the intrinsic brightness in the host galaxy.  

\begin{deluxetable}{cccc}
\center
\tabletypesize{\small}
\setlength{\tabcolsep}{0.02in} 
\tablewidth{0pt}
\tablecaption{The effect of performing background galaxy template subtraction upon our photometry for those SNe in our sample where we have acceptable images.  The Mean, $\sigma$, and Maximum columns represent the mean, standard deviation and maximum deviation in magnitudes between pre and post-template subtraction photometry amongst all observations for all supernovae in our sample that have template data available.}
\tablehead{\colhead{Filter} & \colhead{Mean} & \colhead{$\sigma$} & \colhead{Maximum}}
\startdata
uvw2	&0.11	&0.25	&1.35	\\
uvm2	&0.16	&0.29	&1.38	\\
uvw1	&0.11	&0.20	&1.16	\\
u  		&0.10	&0.21	&1.25	\\
b		&0.08	&0.19	&1.15	\\
v		&0.06	& 0.12	&0.93	\\

\enddata
\\
\end{deluxetable}

\begin{figure}[h!]
\begin{center}
\includegraphics[width=1\textwidth]{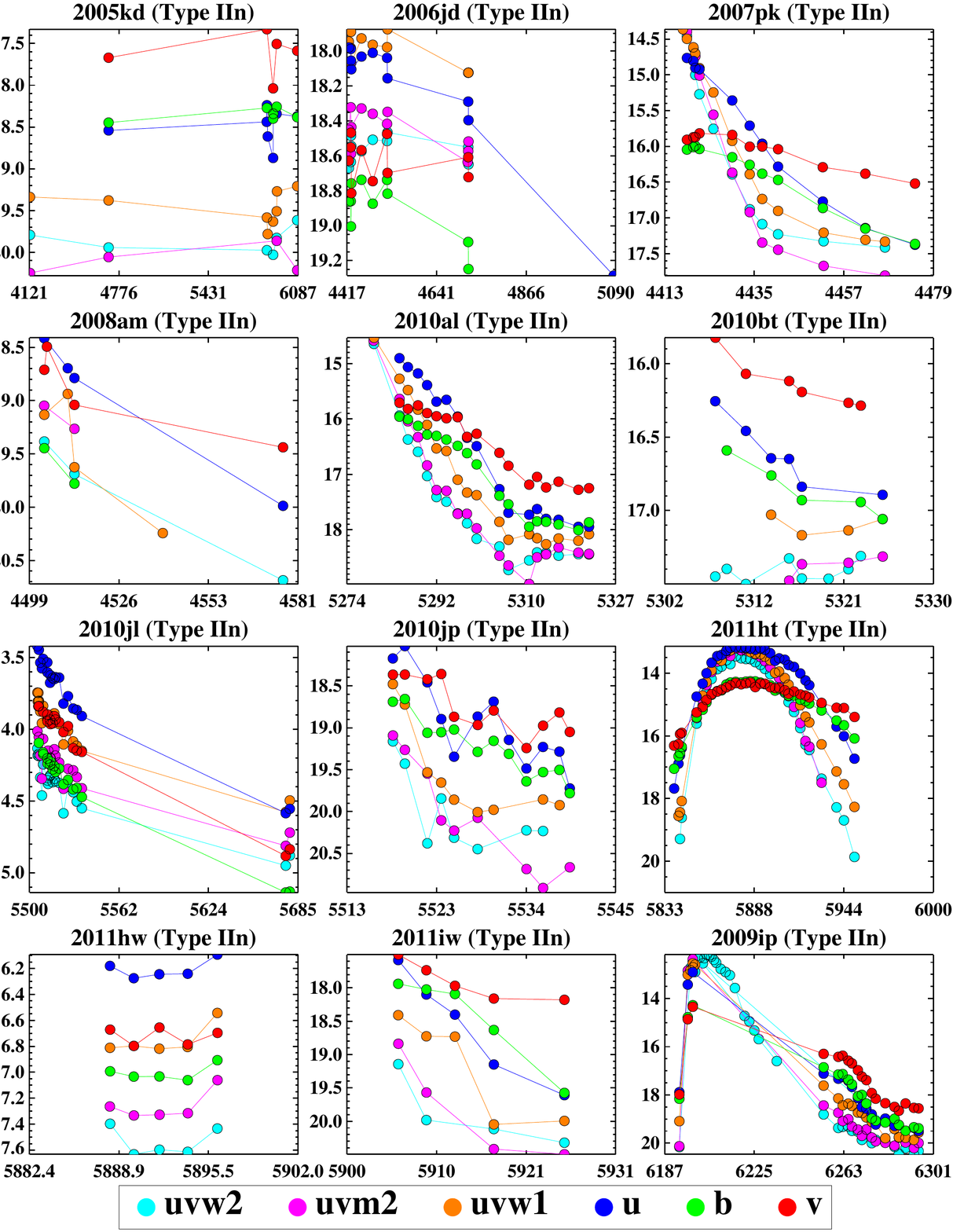}
\caption{\label{indlc_IIn} Individual six filter UVOT light curves for the Type IIn SNe in our sample, arranged by date.  Observations epochs are the Julian Date (JD 2450000+).  }
\end{center}
\end{figure}

\begin{figure}[h!]
\begin{center}
\includegraphics[width=1\textwidth]{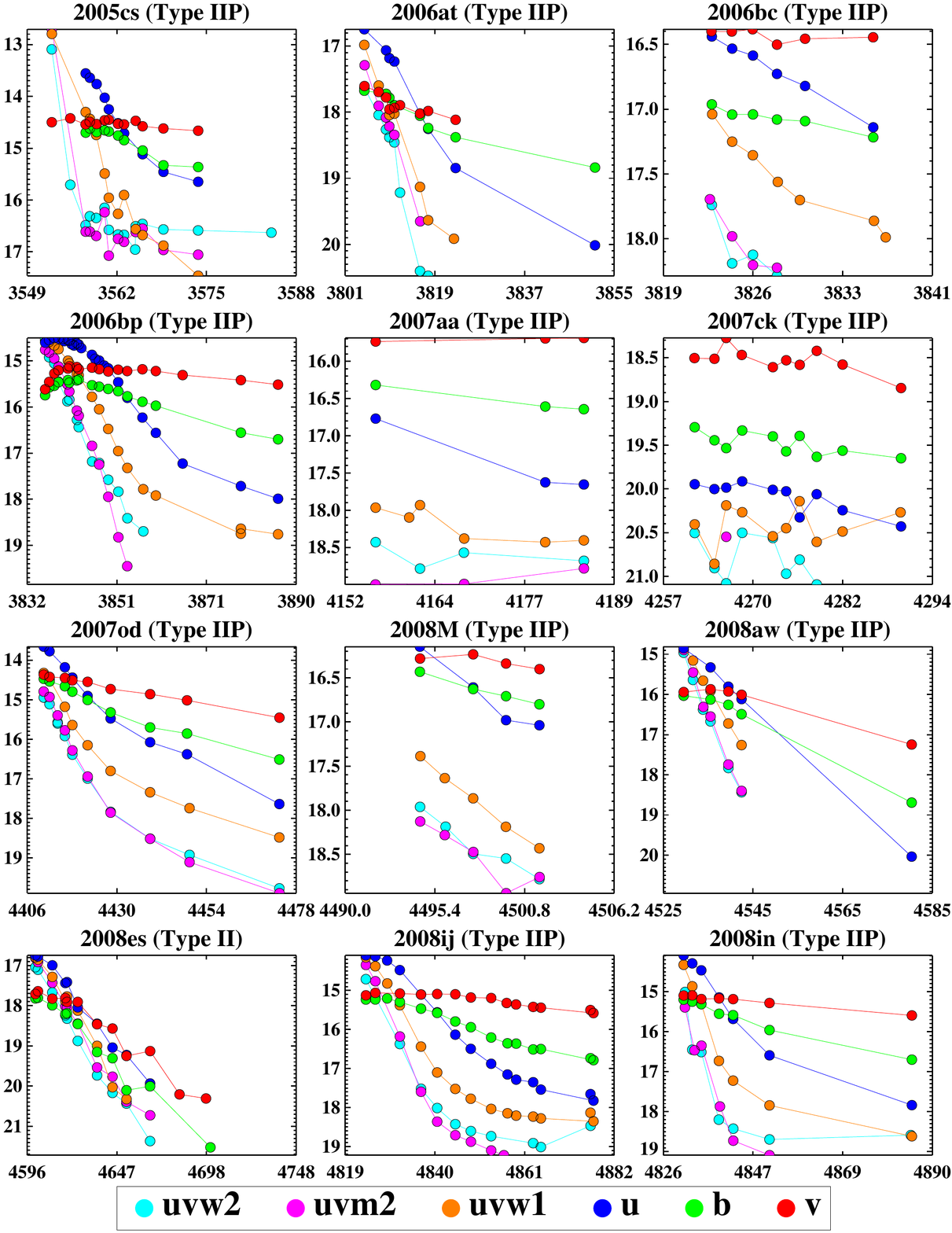}
\caption{\label{indlc_IIpa}  Individual six filter UVOT light curves of the Type II and Type IIP SNe for the years 2005 through 2008 in our sample, arranged by date.  Observations are labeled by shortened Julian Date  (JD 2450000+).}
\end{center}
\end{figure}

\begin{figure}[h!]
\begin{center}
\includegraphics[width=1\textwidth]{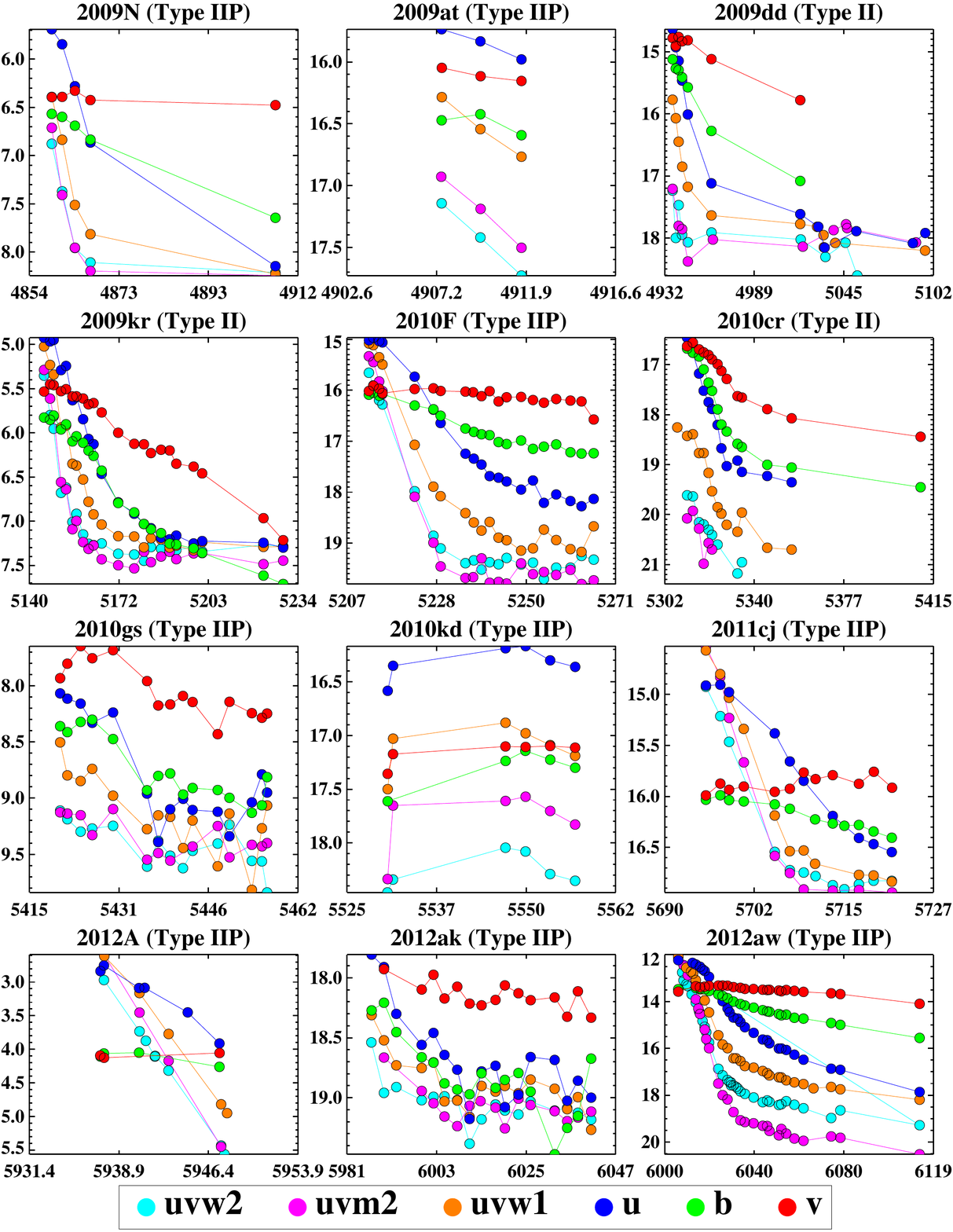}
\caption{\label{indlc_IIpb} Individual six filter UVOT light curves of the Type II and Type IIP SNe for the years 2009 through 2012 in our sample, arranged by date.  Observations are labeled by shortened Julian Date  (JD 2450000+).}
\end{center}
\end{figure}

\begin{figure}[h!]
\begin{center}
\includegraphics[width=1\textwidth]{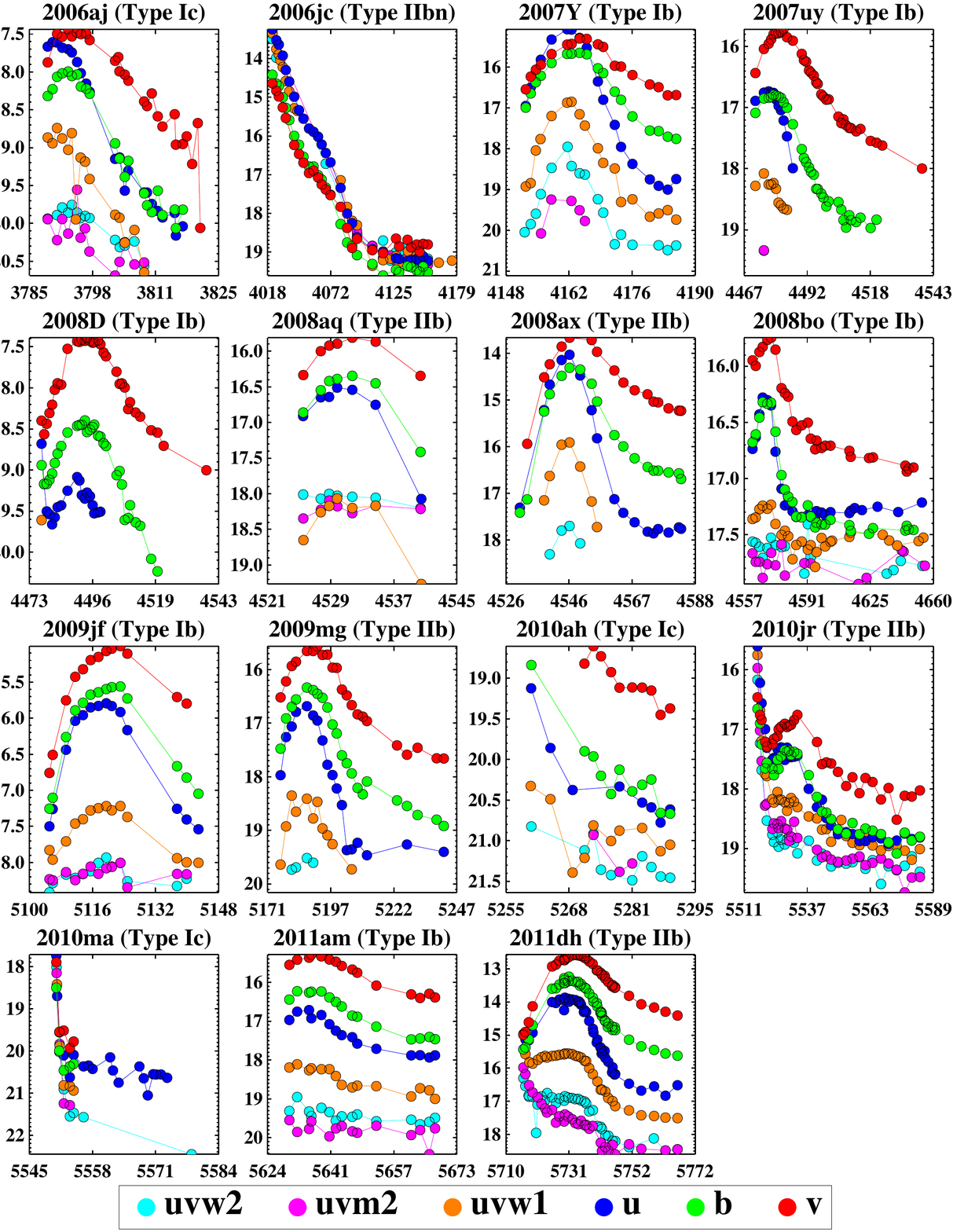}
\caption{\label{indlc_IIbc} Individual six filter UVOT light curves of the Stripped CCSNe in our sample, arranged by date.  Observations are labeled by shortened Julian Date (JD 2450000+).}
\end{center}
\end{figure}

After the extraction of count rates from the $3\arcsec$ aperture, we use a curve of growth model PSF from \citet{Breeveld10}  to perform aperture corrections to a $5\arcsec$ aperture for which the instrument photometry is calibrated.  We use updated Vega zero points from \citet{Breeveld11}, which also contains {\em Swift} AB magnitudes if those are preferred.  Individual six color {\em Swift} UVOT light curves from our sample SNe, broken up by subtype, may be seen in Figures \ref{indlc_IIn} - \ref{indlc_IIbc} for Types IIn, IIP (divided into two plots by year observed), and IIb+Ib/c respectively.  Upper limits and error bars for the individual observations have been omitted for the sake of visibility, however the complete photometry for each object including error bars and upper limits are retrievable at {\em Swift} SNe website: :\url{http://swift.gsfc.nasa.gov/docs/swift/sne/swift_sn.html}.   The median and maximum error bars respectively for our sample in each swift filter are uvw2: 0.14, uvm2: 0.14, uvw1: 0.12, u: 0.1, b: 0.09, v: 0.08 and uvw2: 0.52, uvm2: 0.53, uvw1: 0.52, u: 0.54, b: 0.42, and v: 0.36.

\begin{deluxetable}{cccccccccccccccc}{}\clearpage\rotate
\tabletypesize{\tiny}
\setlength{\tabcolsep}{0.02in} 
\tablewidth{0pt}
\tablecaption{{\small CCSNe Observed By {\em Swift} \label{sne_table}}}
\tablehead{
\colhead{Name} &\colhead{Type}&\colhead{\# UV Obs}&\colhead{RA}&\colhead{Dec}&\colhead{Galactic}&\colhead{Redshift}&\colhead{Distance}&\colhead{$\mu$}&\colhead{Host}&\colhead{Upper Limit}&\colhead{Discovery}&\colhead{Shock Breakout}&\colhead{Template}&\colhead{Bolometric }&\colhead{Ref:}\\
\colhead{} &\colhead{}&\colhead{uvw2 uvm2 uvw1}&\colhead{hr m s}&\colhead{deg m s}&\colhead{E(B-V)}&\colhead{}&\colhead{Mpc}&\colhead{mag}&\colhead{E(B-V)}&\colhead{2450000+}&\colhead{2450000+}&\colhead{2450000+}&\colhead{Image}&\colhead{Light Curve}&\colhead{}}
\startdata
																 	
2005cs	       		&IIP	& 15/ 12/ 12	&13 29 52.78	&+47 10 35.7	&0.031	&0.0015	&8.9		$\pm$0.5	&29.75$\pm$0.12 	&0.01	&		&		&3547.6$\pm$0.5	&C	&Y	&1,2\\
2005kd	       		&IIn	& 7/ 3/ 8   		&04 03 16.88	&+71 43 18.9	&0.233	&0.0150	&63.3	$\pm$4.4	&34.01$\pm$0.14	&0.15	&    		&3683.5	&3686.8	 		&I	&N	&3\\
2006aj$^*$     		&Ic	& 16/ 15/ 16	&03 21 39.71	&+16 52 02.6	&0.126	&0.0331	&145.6	$\pm$9.7	&35.82$\pm$0.14     &0.20	&      		&      		&3784.7	 		&C	&Y	&4, GRB060218\\
2006at	       		&IIP	& 11/ 6/ 8  	&13 12 41.11	&+63 16 45.4	&0.015	&	   	&   				&     				&    		&          	&3802.5	&	 			&C	&N	&5\\
2006bc	       		&IIP	& 4/ 4/ 7   		&07 21 16.50	&-68 59 57.3	&0.181	&0.0045	&22.7	$\pm$4.9	&31.73$\pm$0.50	&0.33	&3811.1	&3819.2	&	 			&C	&Y	&6,7\\
2006bp	       		&IIP	& 14/ 13/ 18	&11 53 55.74	&+52 21 09.4	&0.026	&0.0035	&17.6	$\pm$0.8	&31.23$\pm$0.13     	&0.37	&      		&      		&3833.3$\pm$0.4	&C	&Y	&8,9\\
2006jc	       		&Ibn	& 15/ 61/ 26	&09 17 20.78	&+41 54 32.7	&0.018	&0.0056	&25.8	$\pm$2.6	&32.06$\pm$0.14      &0.03	&4000.3	&4018.3	&4003.0$\pm$50	&C	&Y	&10,11\\
2006jd	       		&IIn	& 10/ 11/ 10	&08 02 07.43	&+00 48 31.5	&0.049	&      		&83.8	$\pm$1.5	&34.62$\pm$0.13 	&0.01	&      		&4021.0	&	 			&C	&Y	&12,13\\
2007Y	       		&Ib	& 16/ 4/ 18 	&03 02 35.92	&-22 53 50.1	&0.019	&0.0046	&18.05	$\pm$1.3	&31.28$\pm$0.16	&0.09	&4119.7	&4147.3	&4145.5$\pm$5.0	&C	&Y	&14,15\\
2007aa	       		&IIP	& 4/ 3/ 6   		&12 00 27.69	&-01 04 51.6	&0.023	&0.0039	&20.5	$\pm$2.6	&31.56$\pm$0.14     &      		&      		&4149.8	&	 			&C	&N	&16,17\\
2007ck	       		&IIP	& 3/ 1/ 7   		&18 23 05.59	&+29 54 01.0	&0.097	&0.0270	&112.51	$\pm$14.	&35.30$\pm$0.31	&    		&      		&4178.5	&	 			&C	&N	&18\\
2007od	       		&IIP	& 11/ 10/ 9	&23 55 48.68	&+18 24 54.8	&0.031	&0.0058	&24.50	$\pm$1.4	&31.91$\pm$0.20	&0.09	&4319.5	&4406.5	&4398.5	 		&C	&Y	&19,20,21\\
2007pk	       		&IIn	& 8/ 9/ 10  	&01 31 47.07	&+33 36 54.1	&0.046	&0.0167	&66.90	$\pm$4.7	&34.13$\pm$0.15	&$<0.13$	&4409.8	&4414.8	&4412.2$\pm$2.0	&C	&Y	&22,23,24\\
2007uy	       		&Ib	& 1/ 1/ 9   		&09 09 35.35	&+33 07 08.9	&0.020	&0.0065	&31.0	$\pm$2.0	&32.46$\pm$0.13 	&    		&4452.5	&4466.1	&	 			&C	&Y	&25\\
2008D	       		&Ib	& 1/ 1/ 1   		&09 09 30.65	&+33 08 20.3	&0.020	&0.0065	&31.0	$\pm$2.0	&32.46$\pm$0.13     	&0.6 		&4474.5	&4475.1	&4474.8	 		&C	&N	&26\\
2008M	       		&IIP	& 5/ 5/ 5   		&06 21 41.28	&-59 43 45.4	&0.040	&0.0076	&40.76	$\pm$8.4	&33.01$\pm$0.48	&    		&4462.5	&4483.5	&	 			&C	&Y	&27\\
2008am$^\dag$	&IIn	& 3/ 2/ 3   		&12 28 36.25	&+15 34 49.0	&0.022	&0.2380	&950.1	$\pm$66.	&39.89$\pm$0.15	&    		&      		&4475.4	&4438.8	 		&C	&N	&28,29\\
2008aq	       		&IIb	& 7/ 7/ 7   		&12 50 30.42	&-10 52 01.4	&0.040	&0.0080	&31.30	$\pm$6.2	&32.45$\pm$0.43	&    		&4506.5	&4523.9	&	 			&C	&Y	&30\\
2008aw	       		&IIP	& 6/ 6/ 6   		&13 04 14.12	&-10 19 12.3	&0.037	&0.0104	&39.12	$\pm$5.7	&32.94$\pm$0.36	&    		&4507.5	&4528.0	&	 			&I	&Y	&31\\
2008ax	       		&IIb	& 4/ 1/ 7   		&12 30 40.80	&+41 38 14.5	&0.019	&0.0019	&8.68	$\pm$1.2	&29.68$\pm$0.31	&0.28	&4528.7	&4529.0	&4528.8	 		&C	&Y	&32,33,34\\
2008bo	       		&Ib	& 16/ 17/ 28	&18 19 54.34	&+74 34 20.9	&0.053	&0.0050	&22.09	$\pm$2.5	&31.71$\pm$0.25	&    		&	      	&4557.5	&	 			&I	&Y	&35\\
2008es$^\dag$	&II	& 10/ 10/ 9 	&11 56 49.13	&+54 27 25.0	&0.010	&0.2100	&848.9	$\pm$62.	&39.64$\pm$0.16	&    		&      		&4582.7	&4574.5$\pm$1.0	&I	&Y	&36,37,38\\
2008ij	       		&IIP	& 11/ 9/ 15 	&18 19 51.81	&+74 33 54.9	&0.053	&0.0050	&22.09	$\pm$2.5	&31.71$\pm$0.25	&    		&4816.5	&4819.9	&4519.2$\pm$2.0	&C 	&Y	&39\\
2008in	       		&IIP	& 7/ 6/ 7   		&12 22 01.77	&+04 28 47.5	&0.020	&0.0052	&13.19	$\pm$1.0	&30.60$\pm$0.20	&0.07	&      		&4827.3	&4825.6$\pm$1.0	&C 	&Y	&40,41\\
2009N	       		&IIP	& 5/ 5/ 5   		&12 31 09.47	&-08 02 56.3	&0.019	&0.0034	&12.60	$\pm$0.9	&30.50$\pm$0.14 	&0.15	&4834.5	&4856.4	&	 			&I	&Y	&42,43\\
2009at	       		&IIP	& 3/ 3/ 3   		&13 46 26.68	&+46 06 09.1	&0.010	&0.0050	&24.15	$\pm$2.7	&31.90$\pm$0.23	&    		&4900.5	&4902.1	&4901.3$\pm$1.0	&C	&Y	&44\\
2009dd	       		&II	& 10/ 10/ 12	&12 05 34.10	&+50 32 18.6	&0.018	&0.0025	&16.24	$\pm$2.1	&31.04$\pm$0.26	&0.43	&4923.5	&4935.5	&	 			&C	&Y	&45,46,47\\
2009jf	       		&Ib 	& 1/ 13/ 9  	&23 04 52.98	&+12 19 59.5	&0.100	&0.0079	&33.85	$\pm$3.0	&32.64$\pm$0.20	&0.03	&5097.5	&5101.8	&5099.7$\pm$2.0	&C	&Y	&48,49\\
2009kr	       		&IIL 	& 22/ 22/ 22	&05 12 03.30	&-15 41 52.2	&0.065	&0.0065	&26.16	$\pm$5.4	&32.03$\pm$0.53	&0.01	&5108.3	&5142.2	&	 			&I	&Y	&50,51\\
2009mg	       		&IIb	& 3/ 1/ 11  	&06 21 44.86	&-59 44 26.0	&0.040	&0.0076	&40.76	$\pm$8.4	&33.01$\pm$0.48	&0.09	&5125.5	&5172.4	&	 			&C	&Y	&52,53\\
2010F	       		&IIP	& 20/ 20/ 20	&10 05 21.05	&-34 13 21.0	&0.095	&0.0093	&32.67	$\pm$8.4	&32.51$\pm$0.57	&    		&5189.5	&5209.8	&	 			&I	&Y	&54\\
2010ah	       		&Ic 	& 10/ 3/ 10 	&11 44 02.99	&+55 41 27.0	&0.011	&0.0498	&208.70	$\pm$14.	&36.60$\pm$0.15	&    		&5246.9	&5251.0	&	 			&I	&Y	&55,56\\
2010al	       		&IIn	& 18/ 18/ 18	&08 14 15.91	&+18 26 18.2	&0.016	&0.0172	&73.40	$\pm$5.1	&34.33$\pm$0.15	&    		&5234.5	&5268.5	&	 			&I	&Y	&57,58\\
2010cr	       		&II 	& 2/ 1/ 7   		&13 29 25.04	&+11 47 46.4	&0.030	&0.0216	&97.50	$\pm$6.8	&34.95$\pm$0.15	&    		&5297.5	&5302.5	&	 			&C	&Y	&59 \\
2010bt	       		&IIn	& 8/ 4/ 4   		&12 48 20.22	&-34 57 16.5	&0.025	&0.0162	&68.70	$\pm$4.8	&34.19$\pm$0.15	&    		&      		&5305.6	&	 			&C	&Y	&61,62\\
2010gs	       		&IIP	& 15/ 4/ 13 	&20 45 39.51	&-05 35 11.0	&0.048	&0.0271	&114.2	$\pm$8.0	&35.29$\pm$0.15	&    		&      		&5410.4	&	 			&U	&Y	&63\\
2010jl	       		&IIn	& 24/ 20/ 24	&09 42 53.33	&+09 29 41.8	&0.024	&0.0107	&48.80	$\pm$3.5	&33.44$\pm$0.15	&0.06	&5479.1	&5503.1	&5480.0$\pm$5.0	&U	&Y	&64,65,66,67\\
2010jp	       		&IIn	& 8/ 9/ 9   		&06 16 30.63	&-21 24 36.0	&0.077	&0.0090	&38.02	$\pm$2.5	&32.90$\pm$0.15	&$<0.16$	&5491.7	&5511.8	&	 			&I	&Y	&68,69\\
2010jr	       		&IIb	& 15/ 19/ 28	&05 19 34.47	&-32 39 14.6	&0.015	&0.0124	&51.30	$\pm$3.6	&33.55$\pm$0.15	&    		&5506.8	&5512.6	&	 			&C	&Y	&70,71\\
2010kd$^\dag$	&IIP	& 6/ 6/ 6   		&12 08 01.11	&+49 13 31.0	&0.021	&0.1000	&414.70	$\pm$29.	&38.09$\pm$0.15	&    		&      		&5515.0	&	 			&I	&Y	&72\\
2010ma$^*$   		&Ic 	& 7/ 5/ 6   		&00 48 55.35	&-34 33 59.5	&0.017	&0.552 	&2096.5	$\pm$100	&41.61$\pm$010 	&$<0.03$	&5549.5	&5550.0	&	 			&I	&Y	&73, GRB101219B\\
2011am	       		&Ib 	& 15/ 14/ 15	&12 16 26.00	&-43 19 20.0	&0.117	&0.0066	&23.675	$\pm$2.2	&31.86$\pm$0.21	&    		&      		&5620.2	&	 			&I	&Y	&74\\
2011cj	       		&IIP	& 12/ 10/ 11	&14 32 53.81	&+11 35 49.3	&0.024	&0.0074	&37.60	$\pm$2.6	&32.88$\pm$0.15	&    		&5686.9	&5690.9	&	 			&I	&Y	&75 \\
2011dh	       		&IIb	& 36/ 33/ 34	&13 30 05.12	&+47 10 10.8	&0.031	&0.0015	&8.03	$\pm$0.7	&29.48$\pm$0.25	&0.0		&5712.6	&5713.4	&5713.0$\pm$0.4	&C	&Y	&76,77,78\\
2011ht	       		&IIn	& 46/ 41/ 48	&10 08 10.59	&+51 50 57.0	&0.009	&0.0036	&19.90	$\pm$0.2	&31.49$\pm$0.25	&0.04	&      		&5833.7	&	 			&U	&Y	&79,80,81,82\\
2011hw	       		&IIn	& 5/ 5/ 5   		&22 26 14.54	&+34 12 59.0	&0.102	&0.0230	&96.20	$\pm$6.5	&34.92$\pm$ 0.14		&    		&      		&5874.3	&	 			&I	&N	&84,85\\
2011iw	       		&IIn	& 4/ 5/ 5   		&23 34 48.20	&+24 45 01.0	&0.049	&0.0230	&93.40	$\pm$6.5	&34.85$\pm$0.15	&    		&      		&5894.6	&	 			&I	&Y	&86\\
2012A	       		&IIP	& 8/ 4/ 5   		&10 25 07.39	&+17 09 14.6	&0.028	&0.0025	&8.10	$\pm$0.2	&29.53$\pm$0.25	&    		&5924.5	&5933.9	&				&I	&Y	&87,88\\
2012ak	       		&IIP	& 22/ 19/ 22	&10 01 27.20	&+36 40 12.0	&0.016	&0.0416	&174.60	$\pm$12.	&36.21$\pm$0.15	&    		&      		&5979.8	&				&I	&Y	&89\\
2012aw	       		&IIP	& 54/ 40/ 46	&10 43 53.76	&+11 40 17.9	&0.024	&0.0026	&10.11	$\pm$0.9	&30.02$\pm$0.20	&    		&6001.8	&6003.4	&6002.6$\pm$0.8	&C	&Y	&90,91\\
2009ip$^\ddagger$	&IIn	& 40/ 20/ 23	&22	23 08.26	&-28 56 52.4	&0.017	&0.0059	&20.4	$\pm$1.6	&31.55$\pm$0.21	&$<0.01$	&		&		&6132.5$\pm$1.0	&I	&Y	&92,93,94,95,96,97\\

\hline
\enddata
\\
$^*$ - Gamma Ray Burst  with Visible Supernova \\
$^\dag$ - Super Luminous Supernova\\
$^\ddagger$ -  2012 explosion, not including 2009\&2010 LBV outbursts \\
{\em References} - 1-\citet{Dessart08};2-\citet{IAUC8555};3-\citet{CBET285};4-\citet{Campana06};5-\citet{IAUC8687};6-\citet{Otsuka12};7-\citet{IAUC8693};8-\citet{Dessart08};9-\citet{IAUC8700};10-\citet{Pastorello07};11-\citet{IAUC8762};12-\citet{Stritzinger12};13-\citet{IAUC8762};14-\citet{IAUC8813},15-\citet{Stritzinger09};16-\citet{IAUC8814};17-\citet{Maguire12};18-\citet{IAUC8843};19-\citet{Andrews10};20-\citet{Inserra11};21-\citet{CBET1116};22-\citet{Pritchard12};23-\citet{CBET1129};24-\citet{Inserra12};25-\citet{IAUC8908};26-\citet{Modjaz09};27-\citet{CBET1214};28-\citet{CBET1262};29-\citet{Chatzopoulos11};30-\citet{CBET1271};31-\citet{CBET1279};32-\citet{CBET1280};33-\citet{Pastorello08};34-\citet{Roming09};35-\citet{CBET1324};36-\citet{CBET1462};37-\citet{Miller09};38-\citet{Gezari09};39-\citet{CBET1626};40-\citet{CBET1636};41-\citet{Roy11};42-\citet{CBET1670};43-\citet{Maguire12};44-\citet{CBET1718};45-\citet{CBET1764};46-\citet{CBET1765};47-\citet{Inserra12};48-\citet{CBET1952};49-\citet{Valenti12};50-\citet{EliasRosa10};51-\citet{CBET2006};52-\citet{Oates12};53-\citet{CBET2071};54-\citet{CBET2125};55-\citet{Corsi11};56-\citet{CBET2198};57-\citet{CBET2207};58-\citet{CBET2220};59-\citet{CBET2281};60-\citet{CBET2269};61-\citet{CBET2250};62-\citet{CBET2252};63-\citet{CBET2397};64-\citet{Stoll11};65-\citet{Smith12};66-\citet{CBET2532};67-\citet{CBET2536};68-\citet{CBET2544};69-\citet{Smith12b};70-\citet{CBET2545};71-\citet{CBET2548};72-\citet{CBET2556};73-\citet{Sparre11};74-\citet{CBET2667};75-\citet{CBET2721};76-\citet{Soderberg12};77-\citet{Arcavi11};78-\citet{ATEL3398};79-\citet{CBET2851};80-\citet{Roming12};81-\citet{Mauerhan12};82-\citet{Humphreys12};83-\citet{CBET2887};84-\citet{CBET2906};85-\citet{Smith12c};86-\citet{CBET2941};87-\citet{CBET2974};88-\citet{CBET2975};89-\citet{CBET3032};90-\citet{CBET3054};91-\citet{Bayless13};92-\citet{ATel4334};93-\citet{ATel4412};94-\citet{Prieto13};95-\citet{Levesque13};96-\citet{Mauerhan13};97-\citet{Pastorello13}

 {\em \# of UV Obs}: The number of observations in each {\em Swift} UV filter (uvw2/uvm2/uvw1) respectively. \\
 {\em Galactic E(B$-$V)}: The MW line of sight reddening in the direction of the SNe  \citep{Schlafly11}.  \\
 {\em Host E(B$-$V)}:  Our adopted SN host galaxy line of sight extinction value (or upper limit) if it has been found in the literature (see references for individual SN).\\
 {\em Upper Limit} The Julian Date(JD) of any reported pre-explosion SN upper limi observations found in the literature.  {\em Discovery}: The JD of the SN discovery image.  Shock Breakout refers to the JD of the SN shock breakout if it has been found in the literature.  \\
 {\em Template}: Have preformed the host galaxy correction described in Section \ref{obs},  C=complete for all filters, I=incomplete for all filters, U=complete for UV filters only.  \\
 {\em Bolometric Light Curve}: Has this object has been included in our Bolometric Light Curve sample (Yes/No)?.  \\
 {\em References}:  Literature found on individual SNe that has been used in this table or paper.  \\
\end{deluxetable}

\section{UV Light Curves of CCSNe}\label{UVLC}
\begin{figure}[h!]
\begin{center}
\includegraphics[width=1\textwidth]{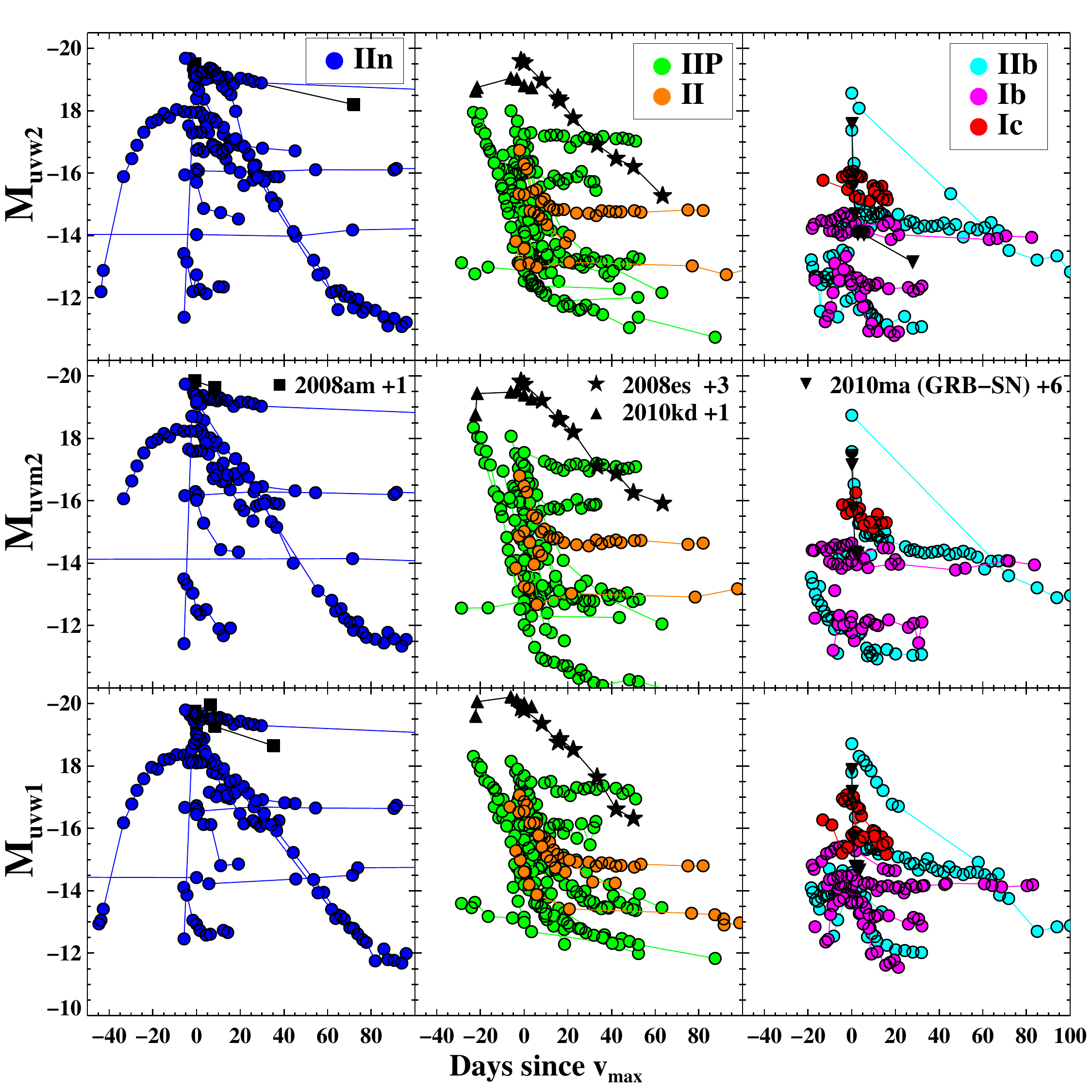}
\caption{\label{absmags}  UV light curves of the CCSNe in our sample,  we plot absolute magnitudes corrected for distance but not dust versus the time since v-band maximum. Several SNe have been shifted vertically to compress the scale - The IIn SN 2008am, two SLSN 2008es and 2010kd, and the GRB-SN 2010ma have been shifted by +1/3/1/6 magnitudes respectively. }
\end{center}
\end{figure}

In Table \ref{sne_table} we present the list of {\em Swift} observed CCSNe used in this table.  Our sample consists of 49 CCSNe and is inclusive of all major subtypes of CCSNe including a number of more exotic SNe such the two 2 GRB-SN 2006aj and 2010ma \citep{Campana06,Starling12}, the Type IIL SN 2009kr \citep{EliasRosa10}, the Type Ibn SN 2006jc \citep{Pastorello07} and several Super Luminous Supernovae \citep[SLSN;2008am, 2008es, and 2010kd;][]{Gal-Yam12}.      The explosion dates of many of these are uncertain, so we will use the {\em v}-band peak time and mag for fiducial purposes to shift our UV light  as seen in Figure \ref{absmags}.  This is the most uniform method available for setting our light curves comparative time scales, but is suboptimal for the SNe cases where {\em Swift} only observes a v-band decline.  Below, we discuss the observed properties of these SNe broken down by subtype.   Dust corrections have not been applied at these wavelengths for Figures \ref{indlc_IIn} - \ref{optabsmags}, as the correction is highly dependent upon both dust model and intrinsic SNe spectrum.  However, Milky Way (MW)  line of sight and Host E(B-V) have been listed in Table \ref{sne_table} as found in the literature.  


\subsection{Type IIn}
The Type IIn SNe in our sample (Figure \ref{absmags}; Top Panes) show the greatest diversity of UV behavior of all our CCSNe subtype samples.  These SNe are often thought to be the product of Luminous Blue Variable (LBVs) stars going SNe, where the expanding SNe ejecta interacts with previous LBV mass loss eruptions (often modeled as a wind, or shell ejections) producing the narrow H$\alpha$ line that characterize this SNe subtype \citep{Chatzopoulos11,Inserra12,Pritchard12,Roming12,Smith12,Smith12c}.  However, there is some disagreement with this progenitor model and it has also been suggested that these could be related to $\eta-$Car type outbursts \citep{Humphreys12}.  In the LBV-progenitor model the observed light curve behavior is produced via a combination of an expanding, cooling hydrogen photosphere driven by the supernovae ejecta and energy injection interaction with the CSM wind/shells.  In terms of the observed light curves, we see a variety of behaviors which may be explained by this physical scenario.  

In some IIn supernovae, such as SN 2007pk and 2010al, the SN peaks quite early in the several days to a week before {\em Swift} observations occur, see Section \ref{uvrise}, and linearly declines across all UV filters.    This is similar to our observations of Type IIP  SNe discussed in Section \ref{IIP}.  This decay in the light curve appears similar to that seen in much of the IIP sample at early times before the optical filters transition into the plateau phase, with an average decay rate of  $\sim 0.27$ mag/day before dropping below {\em Swift} detection limits.   This  is most easily explained by the emission being dominated by the initial SNe ejecta with relatively weak CSM interaction, likely a low-density wind.  In a marked contrast, SN 2011ht has a sharp initial rise of $\sim 6$ magnitudes, followed by a gradual rise to maximum and subsequent decay over  the next $\sim 100$ days, and finally a very rapid decline of several magnitudes in the UV (and more in the optical) at the final observed upper limit.  The differences in behavior of the UV filters are fairly clear in Figure \ref{absmags}, and this behavior is more easily explained by interaction with an optically thick shell.   The rapid increase and decrease in brightness would then occur when the obscured shock begins interacting with or finishes passes through the ejecta shell respectively, and the more gradual rise and fall is moderated by a changing shell and ejecta density/opacity.  In between these two cases we observe a variety of intermediate decay rates with a number of the SNe (e.g. 2005ip and 2006jd)  demonstrating a long lived plateau that we characterize as being driven by energy injection from an optically thin wind or shell.  These plateaus have been observed to have UV magnitudes that may be either brighter or fainter than their optical counterparts, and this is primarily dependent upon the CSM density \citep{Smith09,Stritzinger12} 

\subsection{Type II/IIP}\label{IIP}
The Type IIP SNe in our sample are our most homogeneous subtype.  This tracks with our expectation from the optical light curves as well since this subtype is characterized by $\sim 100$ day optical plateaus whose brightness and duration behave homogeneously throughout the subtype (compared to observed behavior inside of other CCSNe subtypes) and whose variations are thought to be correlated with observables such that they may serve as standardizeable candles \citep{Hamuy02,Dessart08,Kasen09}.  These SNe are thought to have a thick hydrogen envelope which, when ejected, is optically thick and roughly symmetric.  The plateaus are thought to be caused by a  combination of the diffusion of thermal energy from the expanding shockwave into this envelope and a hydrogen recombination wave in the photosphere injecting energy into the ejecta \citep{Chevalier89,Leonard02} after the shock has cooled enough to allow this to occur.  This results in the behavior of the photosphere being well modeled by a dilute blackbody whose properties are primarily driven by photospheric temperature \citep{Dessart05}.  However, in the UV at temperatures below $\sim 7000$ K iron line blanketing is thought to remove or at least diminish this plateau \citep{Kasen09}.  

The IIP UV (and optical) light curves reach maximum very rapidly - thus it is exceptionally rare to catch any UV rise.  {\em Swift} observations taken as early as two days after shock breakout do not detect a clear maximum (see Section \ref{uvrise} and Figure \ref{wellconstr} for more details).  As seen in Figure \ref{absmags}, our light curves  typically begin $0-10$ days before b-band maximum where the plateau phase has yet to begin, and the light curve declines linearly. {
This gradually flattens to a plateau portion by 10 days after {\em v} band maximum in those cases where it is detected.  This suggests that the hydrogen recombination wave does in fact generate a UV plateau in addition to the optical, after the photosphere has expanded and cooled from the initially high temperatures of $\sim 15,000-20, 000$ K  down to $\sim 5,000-7,000$, see \citet{Dessart08,Bersten09,Bayless13} and Section \ref{BBBehave}.  We do however begin seeing significant deviations from dilute blackbody emission here, which is most likely due to iron line blanketing.  This effect is highly temperature and metallicity dependent \citep{Dessart10}, and will tend to absorb a significant portions of the UV spectrum blue-ward of $3500$ \AA\ and transform this into optical and IR emission.   The large observed spread in plateau magnitudes would then be due to a combination of intrinsic explosion energy/$^{56}$Ni (which has been shown to primarily effect the plateau duration, not brightness \citep{Kasen09}), metallically, and dust effects. 

\subsection{Type IIb/Ib/c}
Our stripped core collapse SNe (SCCSNe) come from a diverse range of progenitor systems.  The `typical' IIb/Ib/c is UV-faint with relatively few UV detections \citep{Brown09,Roming09,Oates12}, owing to it's small or nonexistent hydrogen shell..  The UV Light Curves tend to gradually peak and then decline $2-4$ magnitudes below the optical filters, but otherwise trace the optical behaviors.  There are notable exceptions to this rule, however.   Sufficiently early observations of the Type IIb SN 2010jr presented here (See Figure \ref{indlc_IIbc}) have caught the tail of the SN shock breakout cooling phase demonstrating a very early UV bright phase, which may occur in many other SCCSNe if detected early enough, which is caused by the rapid cooling of the SN shock exiting the stellar envelope similar to 2008ax \citep{Roming09}.  We also have two GRB-SNe in this sample.  The GRB adds a power-law component to the SNe spectrum which can both distort the light curve shape and cause the SN to be UV-bright at early times.  Finally, we have the rather unique Ib/c SLSN which, while they are spectrally similar to the typical Ib/c's. are thought to have much more massive progenitors and tend to evolve much more slowly.   

\subsection{UV Rise Time}\label{uvrise}
An interesting apparent behavior of our sample is that while we are typically observing the maximum brightness in the v-band for the SNe shown in Figure \ref{absmags}, we are only seeing UV-maximums in a handful of SNe, most of them SCCSNe - most of our sample IIn and IIPs show no observed maximum,  even those that were observed quite early.  Early observations of the cooling of the SN shock breakout as well as the initial UV rise are driven by SN shock deposition into the progenitor envelope, and are important for improving models of the initial SNe explosion.  The cooling of the shock breakout and subsequent UV rise happens very rapidly, however and so is very hard to observe.  In particular, {\em Swift} SNe observations are dependent upon ground-based survey missions reports of newly discovered SNe - therefore, even with {\em Swift}'s rapid response time, often much less than one day, the earliest we are able to detect SNe is still typically several days after the initial shock breakout.  In addition to this, the shock breakout date is usually determined by either previous upper limits of the same location by the ground based survey, or detailed modeling after the fact - both of which are often unavailable.  In Figure \ref{wellconstr} we present a subset of our larger sample for which we have a relatively well constrained shock breakout date (known to within $\pm 3.5$ days or less).  Using our observations we may then present upper limits on the rise time for this initial UV peak as seen in Table \ref{rt}.  From these it is apparent that  this initial rise happens very rapidly across all of our observed subtypes, in less than $\sim 2-5$ days.  

\begin{figure}[h!]
\begin{center}
\includegraphics[width=1\textwidth]{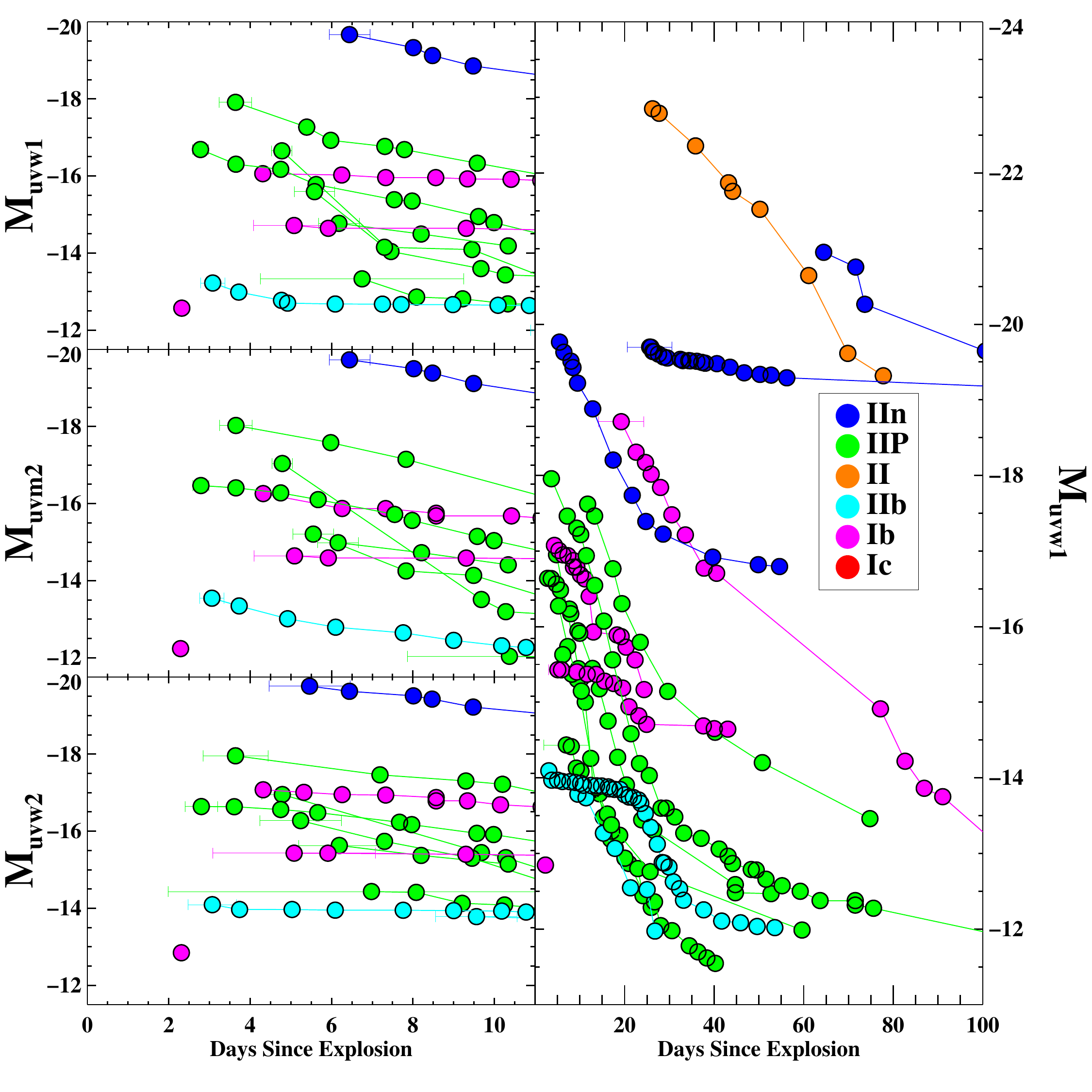}
\caption{\label{wellconstr}  UV Lightcurves of the CCSNe in our sample that have a well constrained rise time, where the x-axis is time since explosion, y-axis is absolute magnitude.  The {\em Left} Panels focus on the  first 10 days after explosion while the right panel are the absolute magnitude uvw1 lightcurves over the first 100 days of {\em Swift} observations. }
\end{center}
\end{figure}

\begin{deluxetable}{cccc}
\center
\tabletypesize{\small}
\setlength{\tabcolsep}{0.02in} 
\tablewidth{0pt}
\tablecaption{Days Since Explosion for the Initial {\em Swift} Observation\label{rt}}
\tablehead{\colhead{Name} &\colhead{uvw2}&\colhead{uvm2}&\colhead{uvw1}}
\startdata
2005cs 	&    4.8 &      4.8 &    4.78\\
2006aj 	&    3.4 &      3.4 &    3.45\\
2006bp 	&    2.8 &      2.8 &    2.79\\
2006jc 	&    19.2 &      19.2 &    19.2\\
2007Y  	&   6.9 &     7.0 &   7.0\\
2007od 	&    11.7 &      11.7 &    11.7\\
2007pk 	&    6.4 &      6.4 &    5.5\\
2008D  	&   2.3 &     2.3 &   2.3\\
2008am 	&    64.4 &      64.4 &    64.4\\
2008ax 	&    1.7 &      11.3 &    1.7\\
2008es 	&    26.2 &      26.2 &    26.2\\
2008ij 	&    47.5 &      47.5 &    47.5\\
2008in 	&    5.2 &      5.5 &    5.23\\
2009at 	&    6.2 &      6.2 &    6.2\\
2009jf 	&   5.0 &      5.1 &   5.0\\
2010jl 	&    25.6 &      25.6 &    25.5\\
2011dh 	&    3.1 &      3.1 &    3.1\\
2012aw 	&    3.6 &      3.6 &    7.2\\
\enddata
\end{deluxetable}

\subsection{Absolute Magnitudes}
Using the data from Table \ref{sne_table}, we convert our observed magnitudes into absolute magnitudes as seen in Figures \ref{absmags} \& \ref{optabsmags}.  In the optical regime we see that for our sample, our peak observed magnitudes range from M$\sim-18$-$-20$ for the IIn's, $-15$-$-18$ for the IIP's, and  finally $-14$-$-18$ for the IIbc's.   In Figure \ref{absmags} our peak UV magnitudes are fairly similar to that of our optical: the IIn's cluster around M$_{uv}=-20$ which is brighter than the IIP's in our sample which are first seen at M$_{UV}=-18$ and our stripped CCSNe at M$_{UV}=-16$.  For the IIn's and especially the IIP's peak brightness appears to occur earlier in time, by up to several weeks.  There are exceptions to this general trend,  and these are the more unique SNe in our sample as detailed previously - the GRB-SNe and SLSN.  The absolute magnitudes in the optical colors behave similarly in our sample (Figure \ref{optabsmags}).  In Particular, we note the substantial Luminosity increase in 2008es, 2010kd and 2010ma - Two SLSN and a GRB supernovae respectively, which have a clear seperation from the rest of the sample.

\begin{figure}[h!]
\begin{center}
\includegraphics[width=1\textwidth]{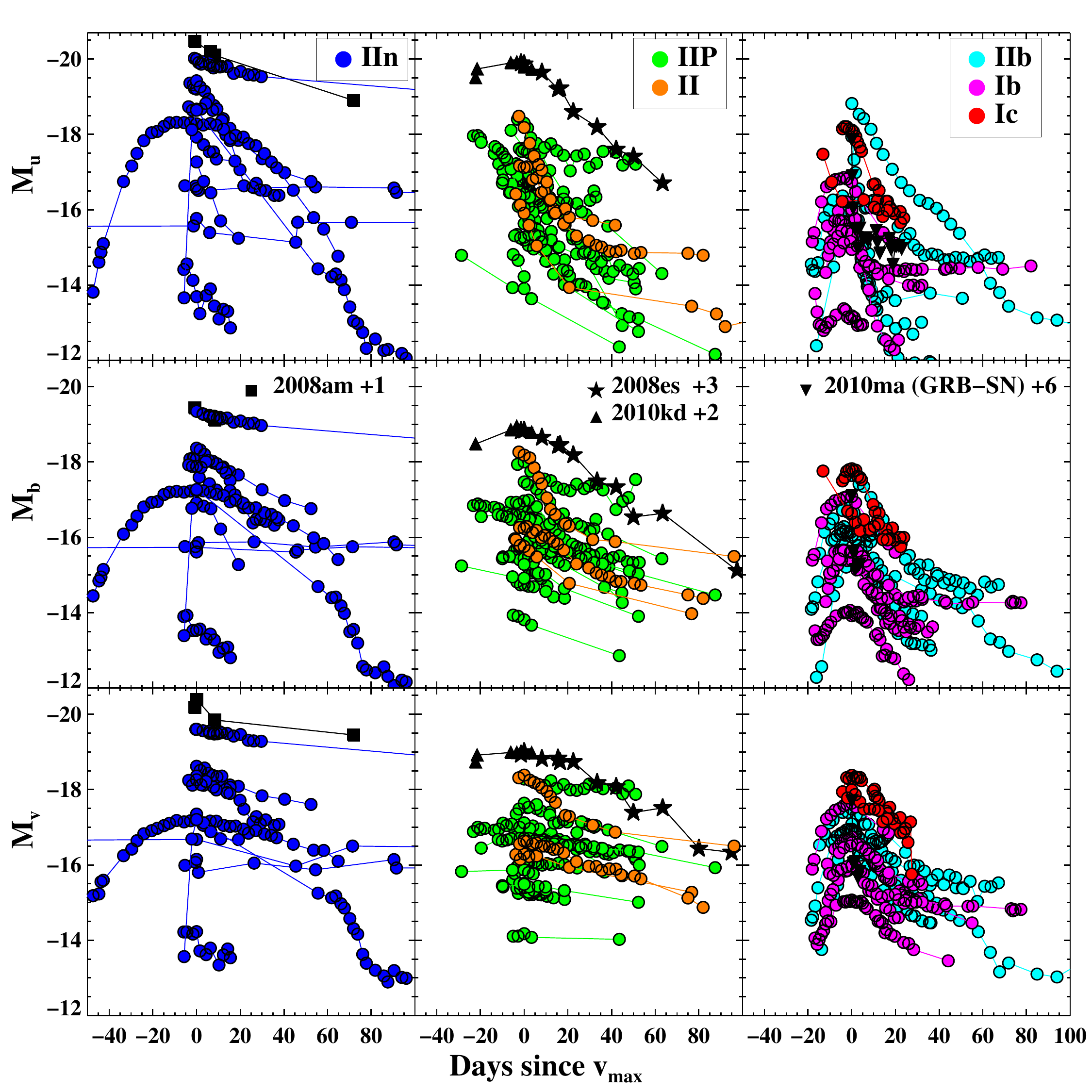}
\caption{\label{optabsmags}  Optical light curves of the CCSNe in our sample,  we plot absolute magnitudes corrected for distance but not dust versus the time since v-band maximum.  Several SNe have been shifted vertically to compress the scale - The IIn SN 2008am, two SLSN 2008es and 2010kd, and the GRB-SN 2010ma have been shifted by +1/3/2/6 magnitudes respectively. }
\end{center}
\end{figure}

\subsection{Color Evolution \& Comparison with Optical}
In Figures \ref{absmags} \& \ref{absmags} we noted that there appear to be several general trends for the IIP and SCCSNe in our sample - that is, the IIP's tend to decline rapidly and enter a UV-plateau phase $10-20$ days after v$_{max}$, while for the SCCSNe, the UV peaks tend to be around v$_{max}$ and have a somewhat flatter light curve shape and evolution than the optical.  To investigate these trends further, we show the UV-v colors for our SNe plotted against time since v-band maximum in Figure \ref{uvtmpl_all}.  Once again, our IIn sample does not have any clear group behavior if the sample is taken as a whole, however we do note that there appear to be a number of SNe with almost flat color evolution between UV$-$v colors of 0 and 2 for days -10 to 40.  This behavior is in contrast with the rest of the sample which indicates a tendency for both rapid increases and decreases in color. \\

\begin{figure}[h!]
\begin{center}
\includegraphics[width=1\textwidth]{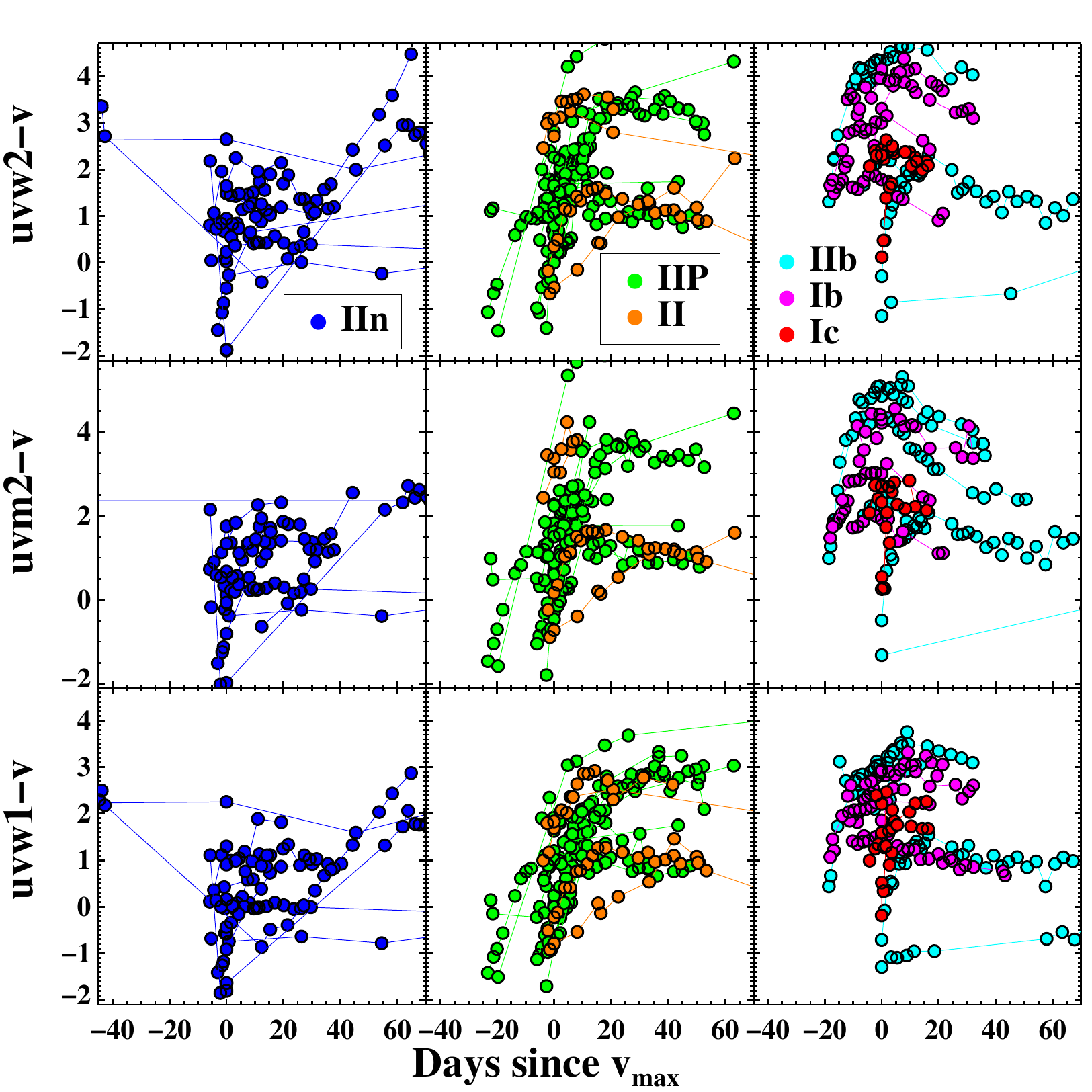}
\caption{\label{uvtmpl_all} {\em Swift}  UV-v color light curves.  The Type II/IIPs appear to have the most homogenous color curves, followed by the stripped core collapse and then The Type IIn.  We note that the outliers in Figures \ref{absmags} \& \ref{optabsmags}  are not apparent on these plots and appear to behave as other SNe of their subtype.   }
\end{center}
\end{figure}

The IIP SNe however appear to have some consensus behavior as you might expect from the more homogenous light curves in Figures \ref{absmags} \& \ref{optabsmags} - all of our IIP SNe appear to start very blue at early times, and then plateau at $\sim 10$ days after v$_{max}$, corresponding to when both the optical v-band light curve and the UV-light curves are in the plateau phase.  Once again the $1-4$ magnitude spread in plateau color should have a lower intrinsic color that is enhanced by differential reddening in the sample which has not been corrected for here, as precise extinction corrections in the UV tend to have large errors due to the significant effect that variations in the $2175$ \AA\ bump cause.  \\

The variety of SCCSNe that {\em Swift} has observed also have some homogenous characteristics.  If we reference Figure \ref{optabsmags}, we note that most of the optical light curves in this sample follow the canonical behavior of a $\sim20$ day rise with all bands peaking at near the same time, followed by a further $\sim20$ day decay that transforms into a slower decline at days $30-60$ as radioactive heating becomes the primary energy source for the SNe.  The UV brightness in these objects is almost always several magnitudes more faint than the optical.  While the UV maximum traces the optical maximum, the overall peak is shallower and less pronounced.  This faint, shallow UV peak compared to the bright, more pronounced optical peak leads to the evident curvature visible in the UV-v colors.


\section{Bolometric Light Curves}\label{BLC}
At early times, a sizable fraction of a CCSNe's bolometric luminosity is in the UV bands.  Using a well observed sub-sample which contains multiple observations in all UV and optical filters as identified in Table \ref{sne_table},  Column 12 we generate bolometric light curves.  We examine the UV characteristics of and contribution to the bolometric light curves as a function of subtype, and derive an empirically based UV corrections for optical bolometric light curves.  We calculate these light curves in the following manner.  Using the \citet{Schlafly11} extinction value for the galactic line of sight extinction component, we generate a range of model blackbodies at different temperatures that have been redshifted to the appropriate value and have galactic extinction applied using the \citet{Cardelli89} analytic model.  If host extinction for the SNe has been determined in the literature we apply another \citet{Cardelli89} model with this value as well.  Otherwise, we fit for this host value as using an upper limit avaliable in the literature if possible, or with an upper limit  of $E(B-V) = 0.3$ if none has been published.  We then perform synthetic photometry upon these model SED's and minimize the $\chi^2$ fit parameter to determine a best-fit model black body temperature and host galaxy reddening where appropriate.  At cooler temperatures much of the SNe flux is red-ward of {\em Swift} UV observations,  while line blanketing starts causing the UV filters to deviate significantly from the blackbody approximation.  This UV deficit caused by line blanketing is  degenerate with the 2175 \AA bump and our fits are of lower quality (see Section \ref{BLC-FC} for further discussion).  We therefore institute a temperature cut of $10,000$ K below which we do not use our fits to calculate the best fit extinction.     Due to the red-leak in the uvw2 and uvw1 filters,  the central wavelengths are not always an accurate representation of the average wavelength from which we are observing the flux, but the process of fitting to these blackbody synthetic magnitudes allows us to model the red leak contribution the observed magnitudes as well as to determine count rate to flux conversion values via the interpolation of these parameters from the blackbody values in \citet{Brown10}.  Using these monochromatic flux densities for each filter we then integrate over the filter bandpass using a trapezoidal integration, careful to avoid filter overlap due to the red leak.  We also compute a bolometric luminosity by applying a Far-UV and Optical/IR correction to our pseudo-bolometric luminosity (which is the integral of the best-fit blackbody at shorter and longer wavelengths that the UVOT bandpass) for the Far-UV and Optical+IR corrections respectively.  We use updated UV filter curves from \citet{Breeveld10} which have a modified red-leak shape from the initial curves depicted in \citet{Poole08}.   The SNe 2008ax, 2009mg, 2010cr, and 2011am all have few uvm2 detections but numerous uvw2 and uvw1 detections.  In these particular cases we did not use the uvm2 filter in the previously described calculations.  Bolometric light curves calculated here are available in Machine Readable format at the same location as the UVOT photometry files referenced in Section \ref{obs}. 
 
 
\subsection{Bolometric Light Curve Flux Completeness and Accuracy}\label{BLC-FC}
In using Blackbody functions to assist in our handling of the {\em Swift} UV-filters red-leaks and calculation of the SNe bolometric light curves, we have introduced some model dependence into these calculations.  First, we may ask how well we are fitting our results. To examine this we look at the residuals between synthetic magnitudes from our best fit blackbody for our sample in Figure \ref{spectint} ({\em Left}).  We see that most of our calculations have reasonable residuals compared to our median and maximum photometric errors quoted in Section \ref{obs}, and at more than 95\% of our epochs our model photometry fits to within the observed 2-$\sigma$ photometric errors of the observations, although there are a number of worse fits that suggest that either the observed datapoint is inaccurate (one filter deviating) or that the model is improbable (large errors in several filters).    To quantify how accurately these calculations are reproducing observed results we take a number of HST UV spectra combined with ground based Optical spectra of CCSNe (1993J, 1994I, 1998S, 1999em) as well as hydrodynamical models of the Type IIP SNe 2005cs and 2006bp from \citet{Dessart08} and generate synthetic magnitudes in the UVOT bands.  For 2005cs and 2006bp we used the known values for host reddening listed in Table \ref{sne_table}, while for the other SNe we fit for extinction.  We then run these `observations' through our bolometric light curve pipeline and compare our calculated pseudo-bolometric measurement ({\em Swift} observed bands only, $1600-6000$ \AA). with the integrated flux directly from the observed spectra or models.  The results from this may be seen in Figure \ref{spectint}.  We reproduce these observed values to better than 7\% at Temperatures ranging between 5000 and 30000K.  Below 5000K this deviation grows as we appear to increasingly underestimate the intrinsic flux as the blackbody peak is red-ward of the {\em Swift} bands. The difference in flux is due to a combination of both error in the fit between the blackbody and the underlying spectral continuum as well as spectral lines/deviations from a blackbody.  

\begin{figure}[h!]
\begin{center}
\includegraphics[width=.49\textwidth]{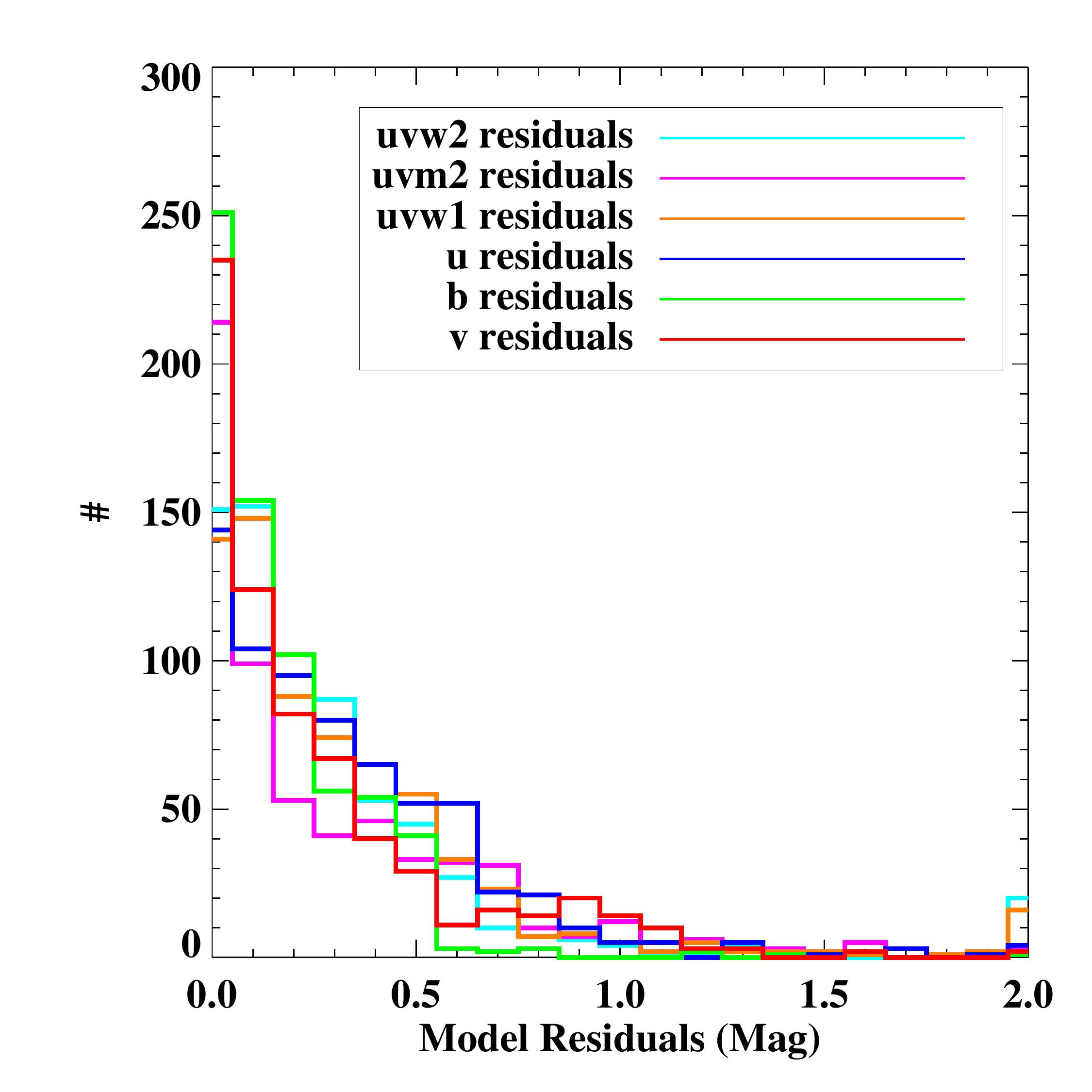}
\includegraphics[width=.49\textwidth]{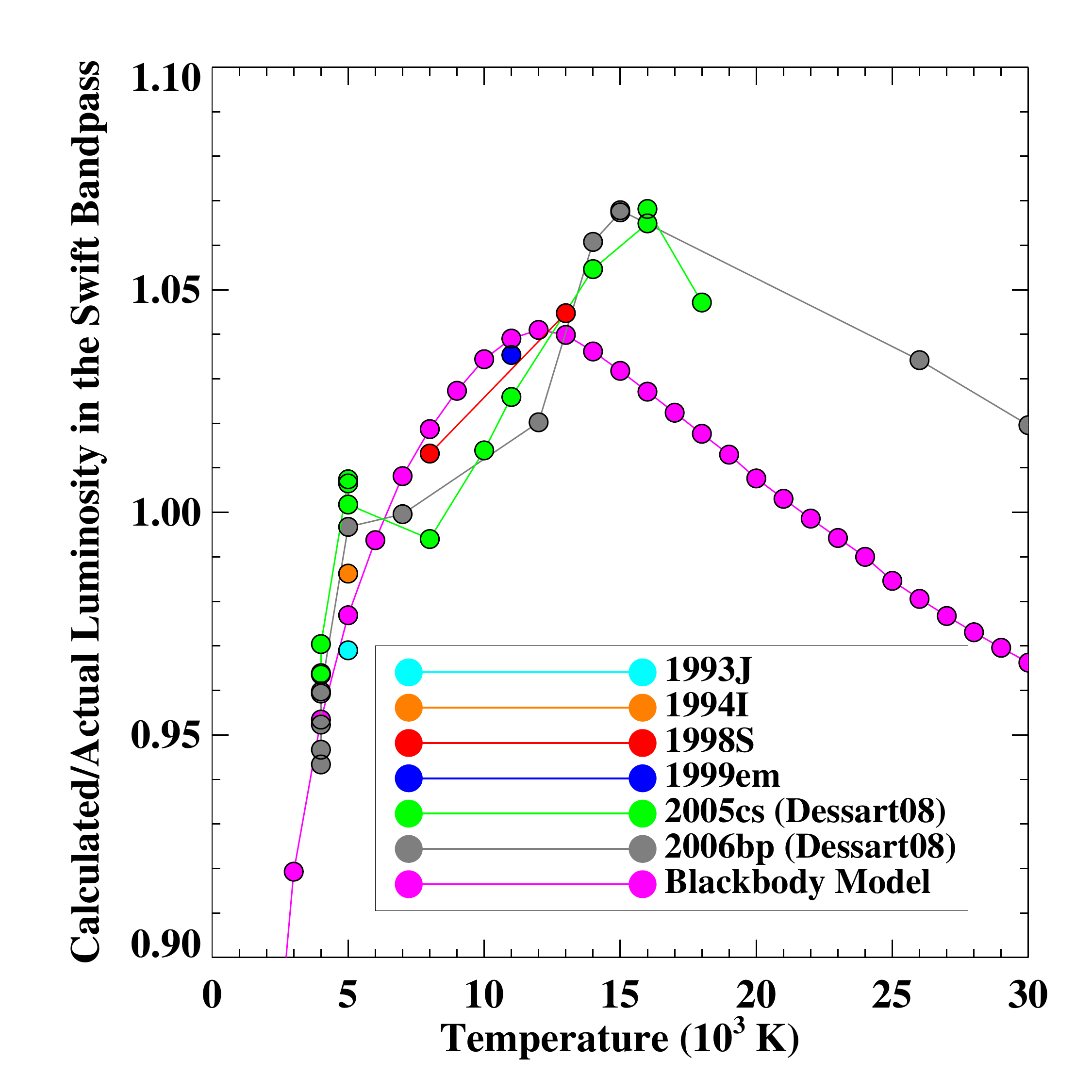}
\caption{\label{spectint} {\em Left}: A histogram of our model residuals (i.e. - the difference between synthetic magnitudes for our best-fit blackbody model and the observed filter values) per filter for every observation in our sample.  {\em Right}: A comparison of our calculated peudo-bolometric flux from observed/model spectra and the intrinsic values.  }
\end{center}
\end{figure}

As the primary motivation of these bolometric light curves is to analyze the UV flux contribution to the bolometric luminosity of observed SNe, we must first ask how well we are sampling these SNe with the {\em Swift} bandpass.  Our observations have a high UV-completeness.  In Figure \ref{corrfrac} ({\em Left}) we show our interpolated UV correction for flux that originates blue-ward of the observed {\em Swift} filters as a function of optical color.  What we see is that in all but the bluest of observations we are below a 10\% UV correction factor, and all observations are below a 30\% value.  This is a reduction in missed blue flux by a factor of 2.5 - 6 in comparison with what may be done on the ground (c.f. \citet{Bersten09}).  Unfortunately, UVOT's reddest filter is the v-band which terminates at 6000 \AA.  This means that we lose a significant portion of the flux as the SNe spectra cool and redden.  When the SNe are UV bright our IR correction may be low at 10-20\% (Figure \ref{corrfrac}, {\em Right}), however this increases as the bulk of the flux shifts red-ward of the optical.  In UVOT's worst cases, we sample only 5-10\% of the flux for observations of red SNe (primarily SCCSNe) at late times.  We may combine these two observations to instead look at the {\em Swift} observed fraction of SNe light as a function of time since the SNe explosion.  We find that when we catch these objects early we have a high total flux completeness value as most of the SNe flux is in the UV.  By days $\sim30-40$ the UV brightness has decreased substantially and we are left with primarily optical and IR flux where only $\sim20$\%  (or even less in a few rare cases) is in the {\em Swift} photometry bands (Figure \ref{optvtime}).  The Type IIn SNe in our sample appear to deviate from this slightly and have a much longer interaction lifetime due to the CSM interaction with the SNe shock which helps to keep light curves UV bright even at late times.  The SLSN appear to also behave on a longer timescales, keeping a high flux completeness at late times.   {\em The addition of ground based red-optical and IR data is necessary to bring these observations up to near flux completeness at these later times for most of these SNe.}   A follow-up paper is in progress where we perform a similar analysis on a subsample of these objects while incorporating comprehensive ground based observations.    

\begin{figure}[h!]
\begin{center}
\includegraphics[width=.49\textwidth]{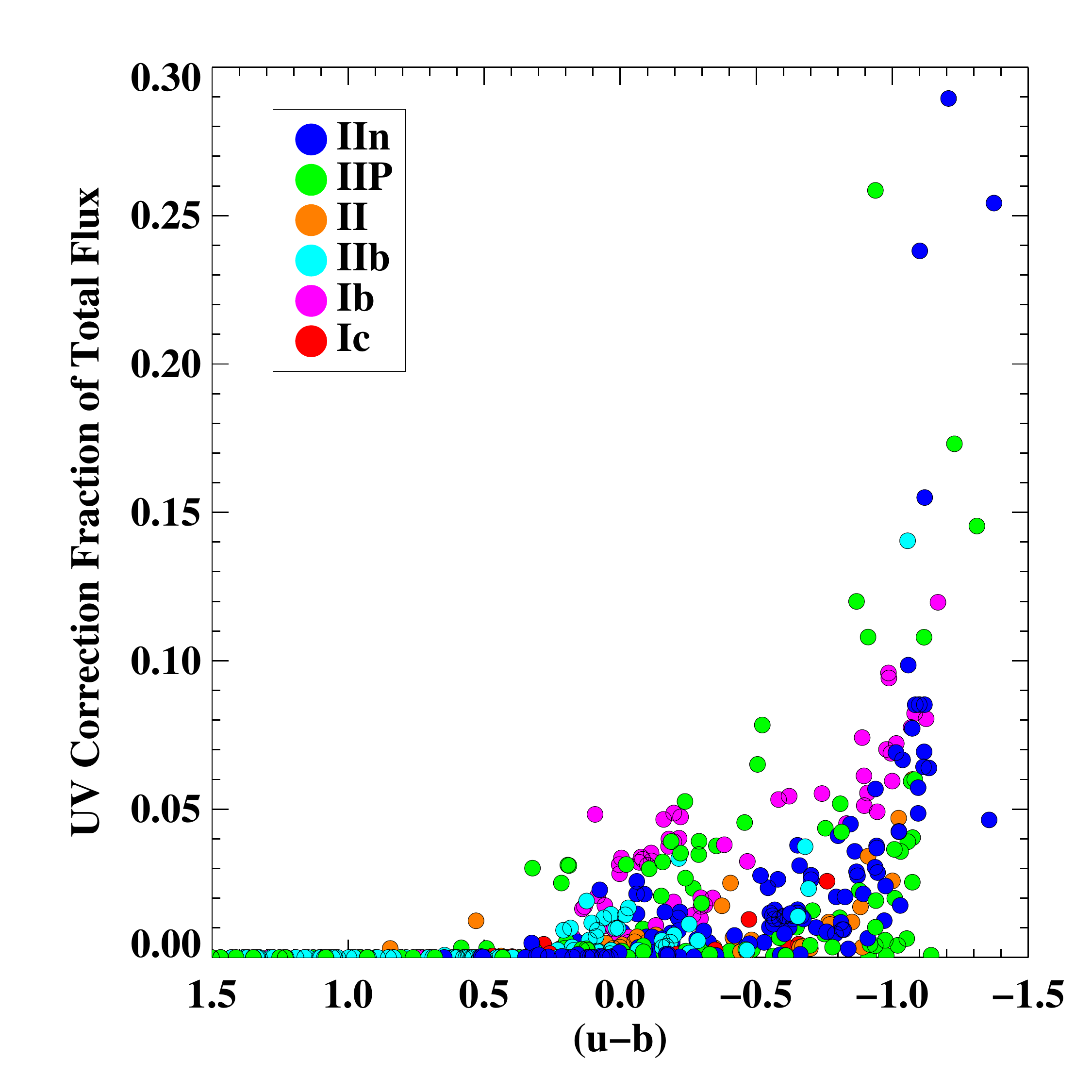}
\includegraphics[width=0.49\textwidth]{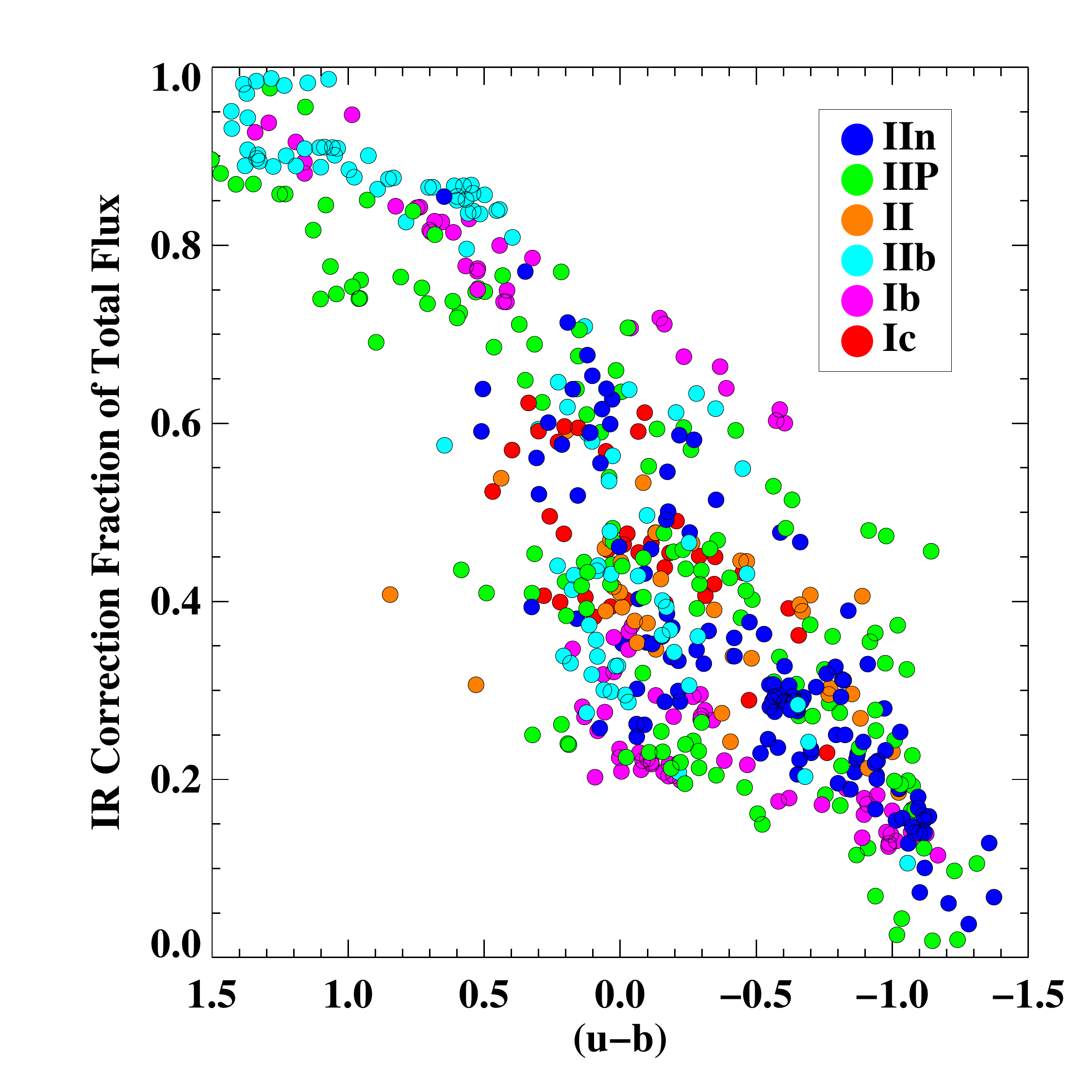}
\caption{\label{corrfrac}  {\em Swift} UV ({\em Left}) and IR ({\em Right}) correction factors as a function of observed colors.}
\end{center}
\end{figure}

\begin{figure}[h!]
\begin{center}
\includegraphics[width=.49\textheight]{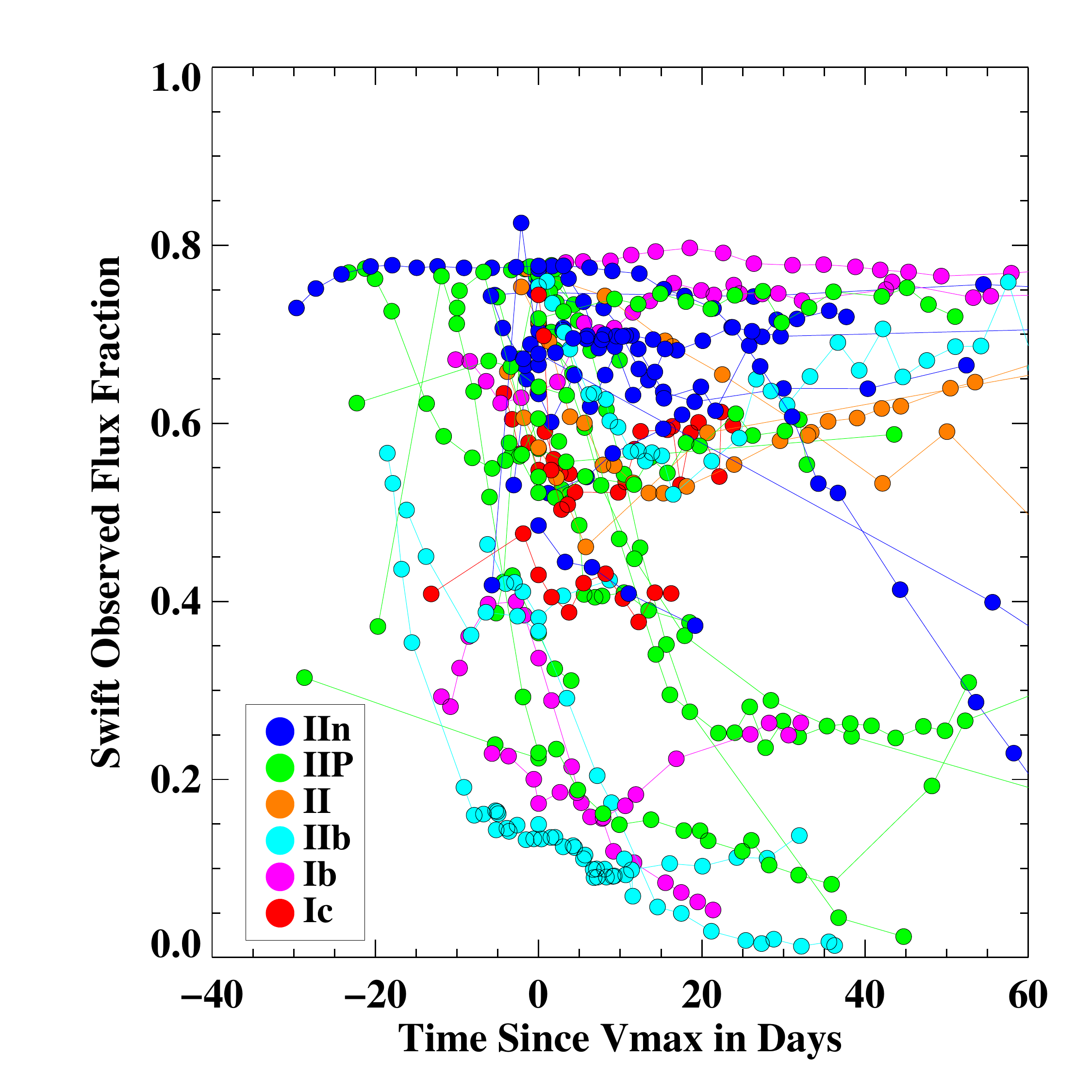}
\caption{\label{optvtime}  The observed fraction of the SNe bolometric flux versus time since v-band maximum.   As the CSM interaction drives the bolometric luminosity the IIn's in our sample exhibit a large UV flux at much later times than the other CCSNe in our sample.  Indeed, for most other SNe in the UVOT observed flux decreases to about 20\% sometime between days  20 and 40 while IIns remain both more variable and more UV bright at later times.}
\end{center}
\end{figure}

While modeling the spectra of a SNe as a dilute blackbody has long been used as a first order approximation, the presence of metal lines in the UV spectra is expected to deviate from a blackbody. These lines, in addition to residual errors from the extinction correction and the uncertainty in fit at some epochs due to the limited flux in the {\em Swift} bandpass are expected to generate some appreciable error.  To quantify this, we use hydrodynamic models of the two Type IIP SNe 2005cs and 2006bp as presented in \citet{Dessart08}.  Using these UV-optical model spectra which have well defined photospheric temperatures we generate synthetic {\em Swift} magnitudes and run these through our fitting algorithms to examine how our measurements compare to the model parameters.  We find that at temperatures that are hotter than about 9,000 K, our measured temperatures are systematically biased by about 20\% cooler than the model's photospheric temperature, while below 9,000 K the photospheric temperature tends to be $\sim40$\% hotter.  For temperatures above 8,000 K we attribute this bias primarily to the depressed model flux compared to the blackbody values lowering the best-fit temperatures.  At the lower temperatures our flux-completeness becomes rather low as little of the flux is in the UV and the primary bias there is due to the high uncertainties in the UV observations and fitting.  This may be seen in Figure \ref{bbdev} ({\em Right}).  

\begin{figure}[h!]
\begin{center}
\includegraphics[width=.49\textwidth]{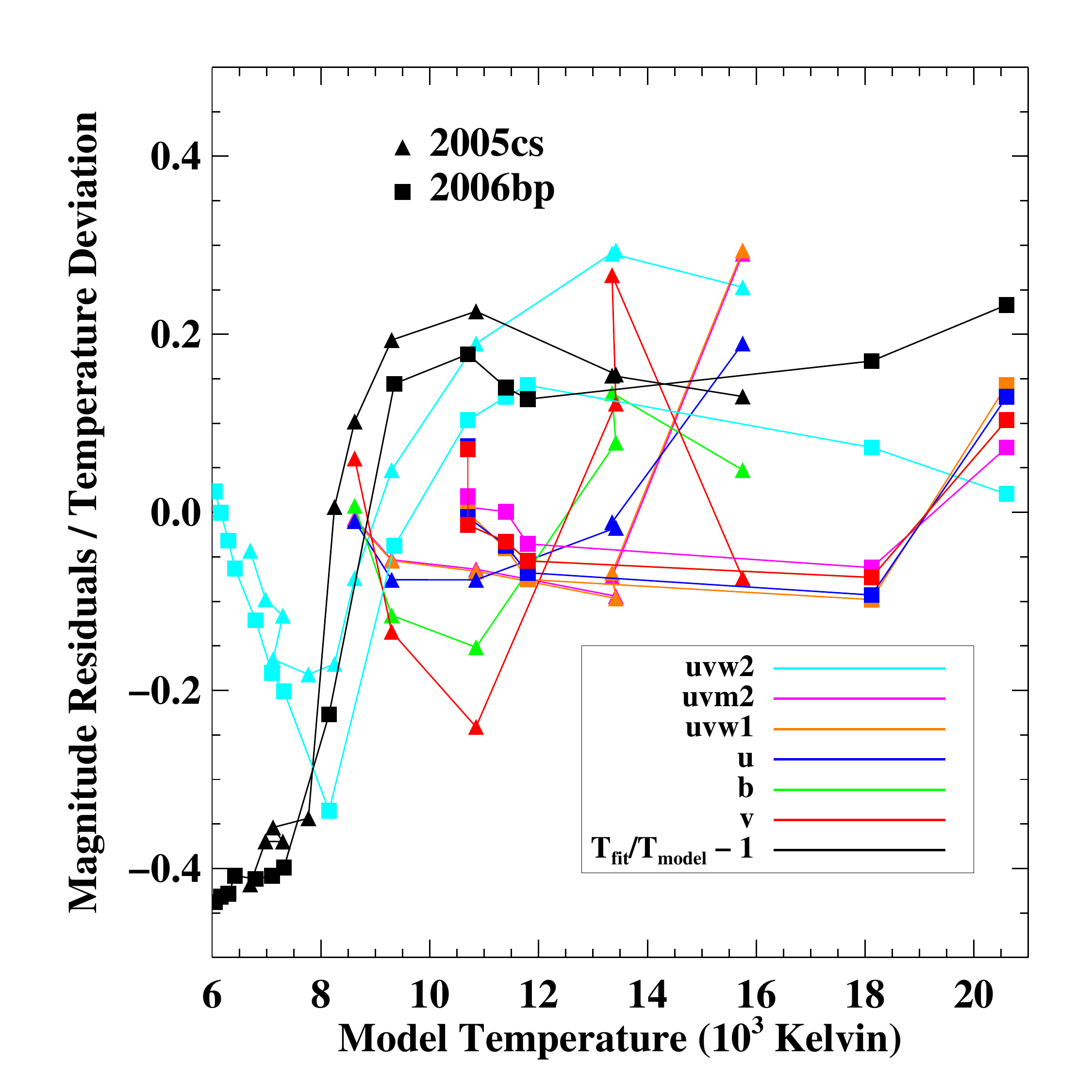}
\includegraphics[width=.49\textwidth]{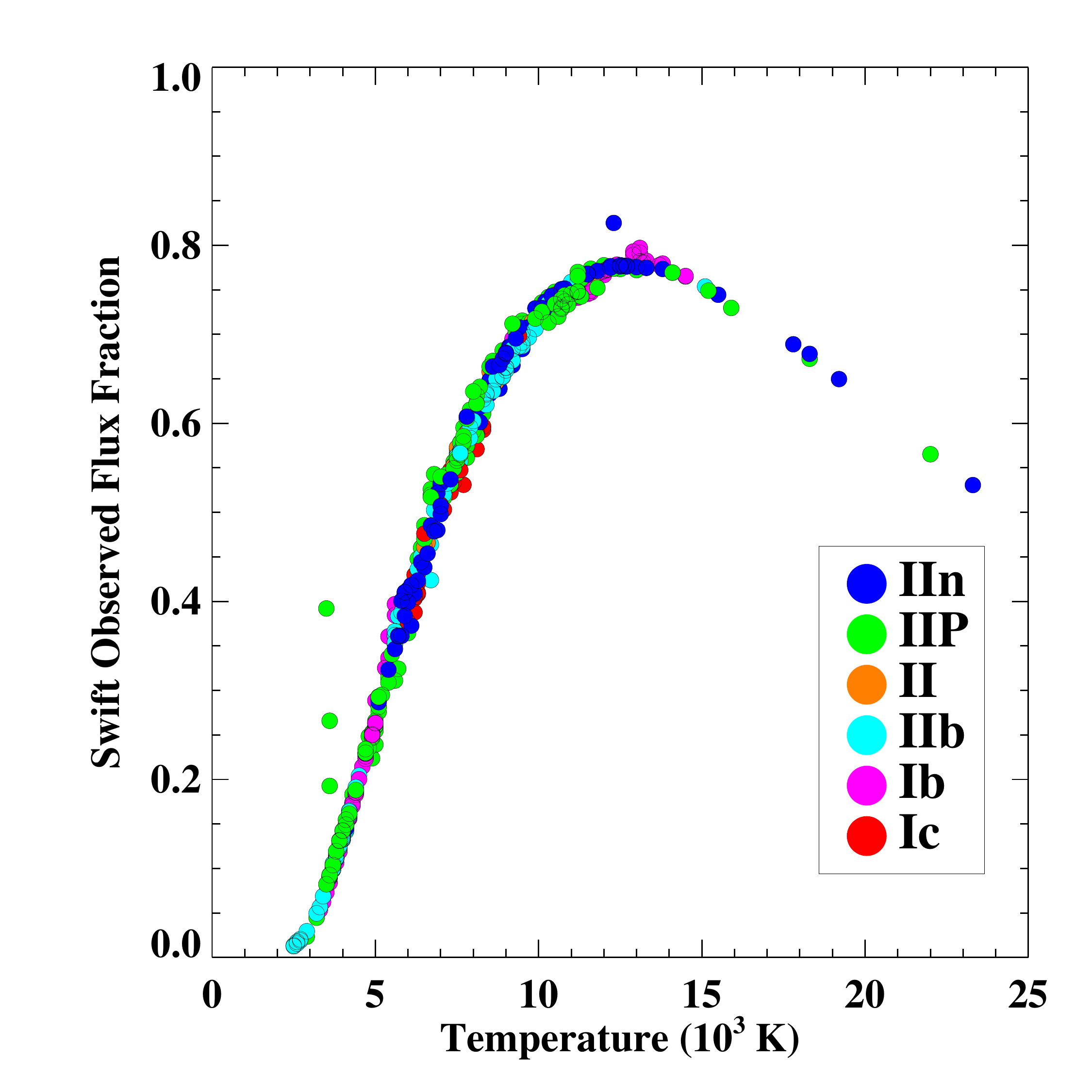}
\caption{\label{bbdev} {\em Left}: A comparison between best-fit Temperatures to models of 05cs and 06bp from \citet{Dessart08} and the models actual photospheric temperatures, as well as magnitude residuals between the models' synthetic magnitudes and our best fit magnitudes. {\em Swift} Blackbody parameters from our best fits.  {\em Right}: {\em Swift} observed bolometric flux fraction as a function of temperature for our entire SNe sample. Each data point represents the calculated observed flux fraction for an individual observation of a particular SNe, color coded by observed SNe subtype.}
\end{center}
\end{figure}

\subsection{Bolometric and UV Corrections}\label{UV-C}
For SNe that lack IR and UV observations, it can be convenient to define a bolometric correction value, ie a value that transforms an observed optical V-band value into a bolometric magnitude empirically using a different observed sample.  While we lack IR observations in this dataset, many of our observations are early enough that this is not a significant handicap, and we are able to calculate this conversion as, 
\begin{equation}\label{bce}
BC=m_{bol}-[V-A_V]
\end{equation}
where BC is the bolometric correction, $m_{bol}$ is the total bolometric magnitude, V is the observed v-band magnitude, and A$_V$ is the visual extinction.  Bolometric Corrections for the SNe in this sample may be seen in  Figure \ref{bolcorr}.  We calculate polynomial fits to this data, which are listed in Equations \ref{bcu} \& \ref{bcb}.  

\begin{figure}[h!]
\begin{center}
\includegraphics[width=.49\textwidth]{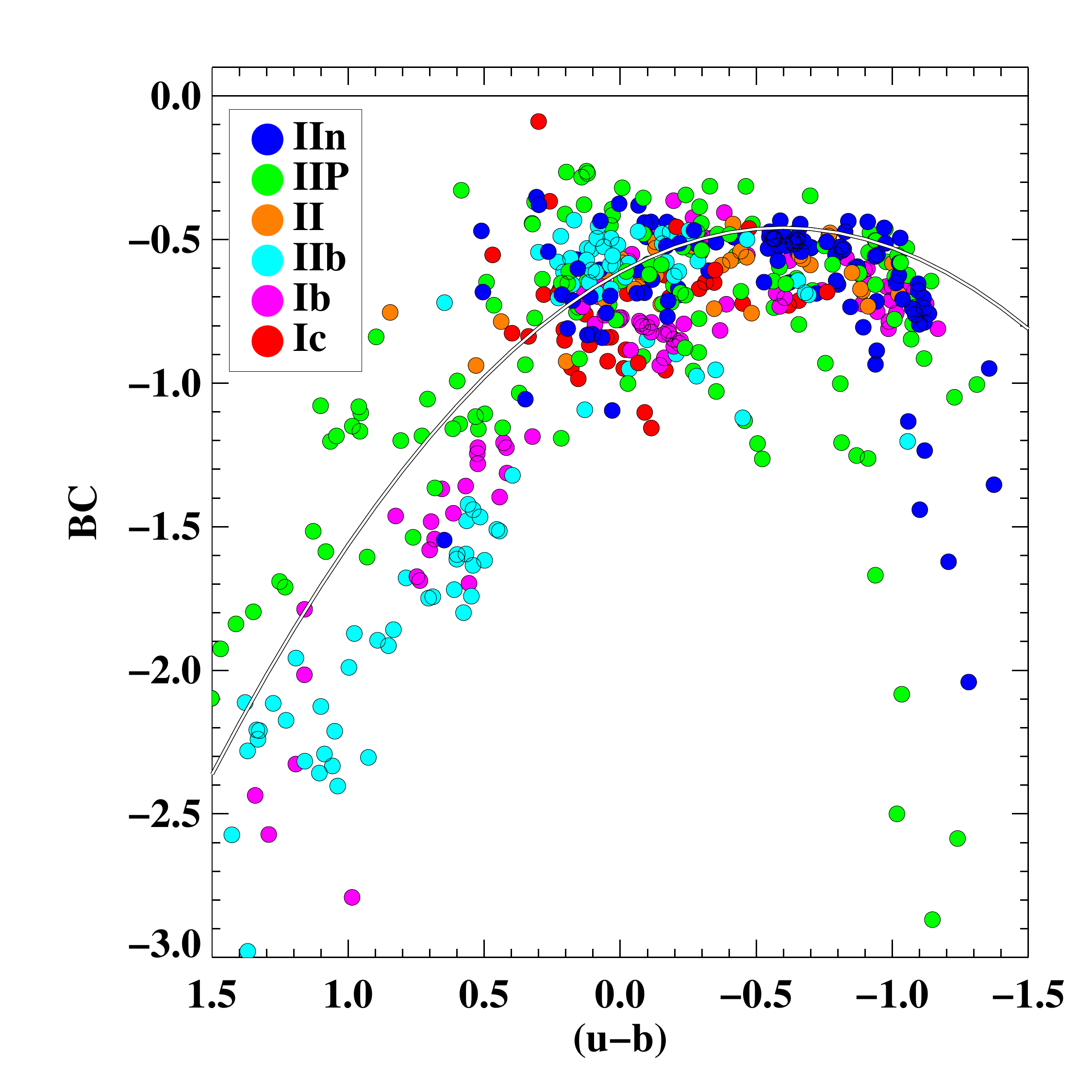}
\includegraphics[width=0.49\textwidth]{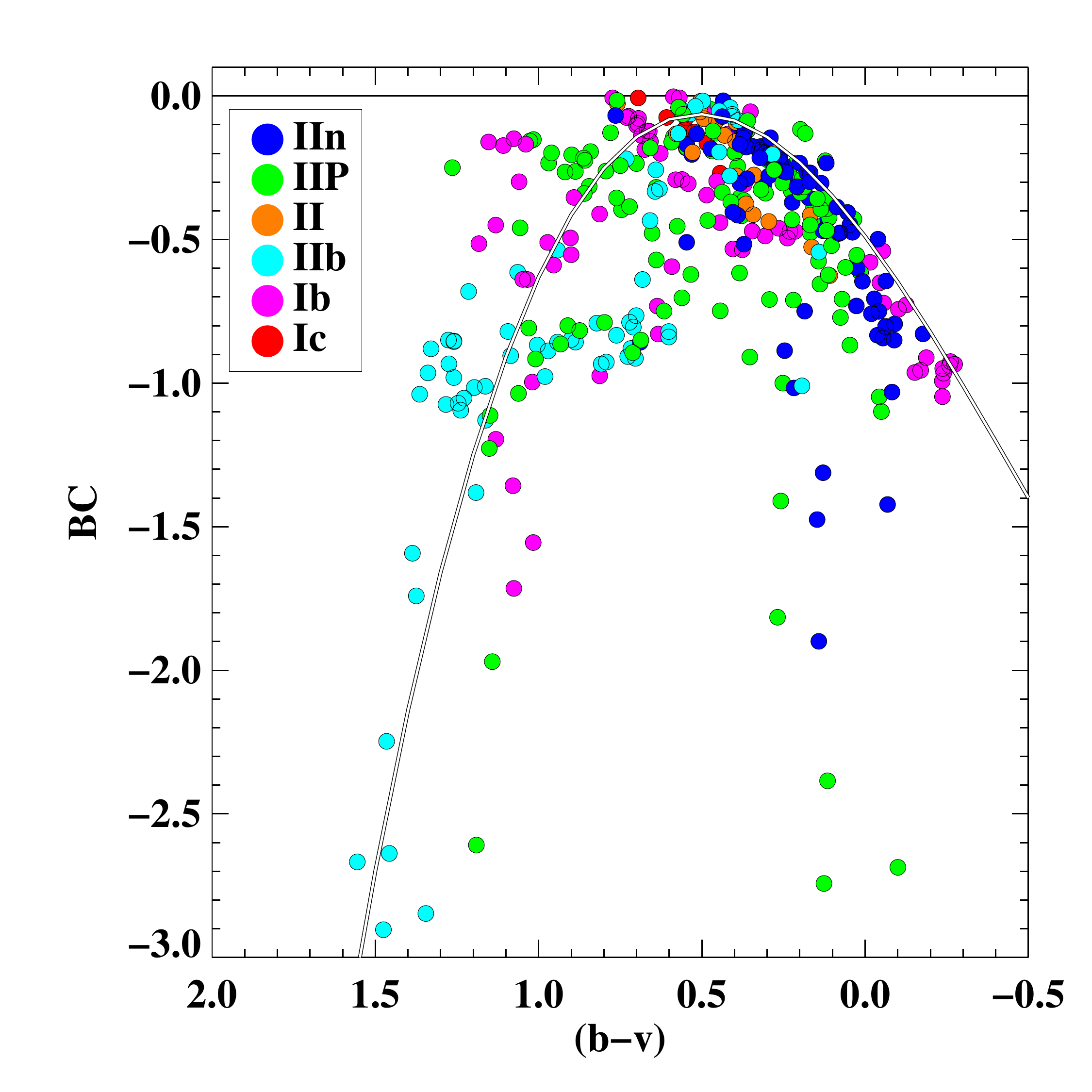}
\caption{\label{bolcorr}  The calculated Bolometric Correction as a function of optical colors.  The $u$ band here  from {\em Swift} and has a bluer cut off than ground based filters.  The gray lines represent the Equations \ref{bcu} ({\em Left} Panel) \& \ref{bcb} ({\em Right} Panel) which are polynomial fits to the observed data discussed in Section \ref{UV-C}}
\end{center}
\end{figure}

\begin{equation}\label{bcu}
BC (u-b) = -0.6133-0.517\times (u-b)-0.4326\times (u-b)^2\\\
\end{equation}

\begin{equation}\label{bcb}
BC (b-v) = -0.4888-1.5046\times (b-v)-0.9697\times (b-v)^2-0.6768\times (b-v)^3\\\
\end{equation}

While this is useful, the proliferation of ground based IR transient telescopes means  that rather than focusing on a total bolometric correction, we should perhaps leverage {\em Swift}'s unique strengths and instead give a total  UV correction, where we supply a magnitude correction for the SNe flux blue-ward of b-band.
This is magnitude value calculated similarly to the Bolometric Correction (Equation \ref{bce} discussed above, but $m_{bol}$ is instead $m_{uv}$) and as such is a distance independent value.  In Figure \ref{colorcorr} we plot this value versus u$-$b colors ({\em Left}) and b$-$v colors {\em right}.  

We include the u$-$b colors  for use with {\em Swift} UVOT observations as an estimate for when UV filters might be lacking.  Care should be taken not to use these with the more common Johnson U filter, as the {\em Swift} filter has a cutoff blue-ward of the Johnson U-band this can not be applied to ground based data.  This does illustrate the point that the space u-band is a much more efficient tracer of the UV flux than the other {\em Swift} optical filters, since it is both closer in wavelength and similarly effected by spectral effects such as line blanketing.  We perform a linear fit for $u-b$ and $b-v$ respectively, with the best-fit values listen in Equations \ref{uvcfu} and \ref{uvcfb} respectively.  The standard deviation of the data about these fits are $\sigma=0.34$ and 0.63 for u$-$b and b$-$v respectively.  

\begin{figure}[h!]
\begin{center}
\includegraphics[width=.49\textwidth]{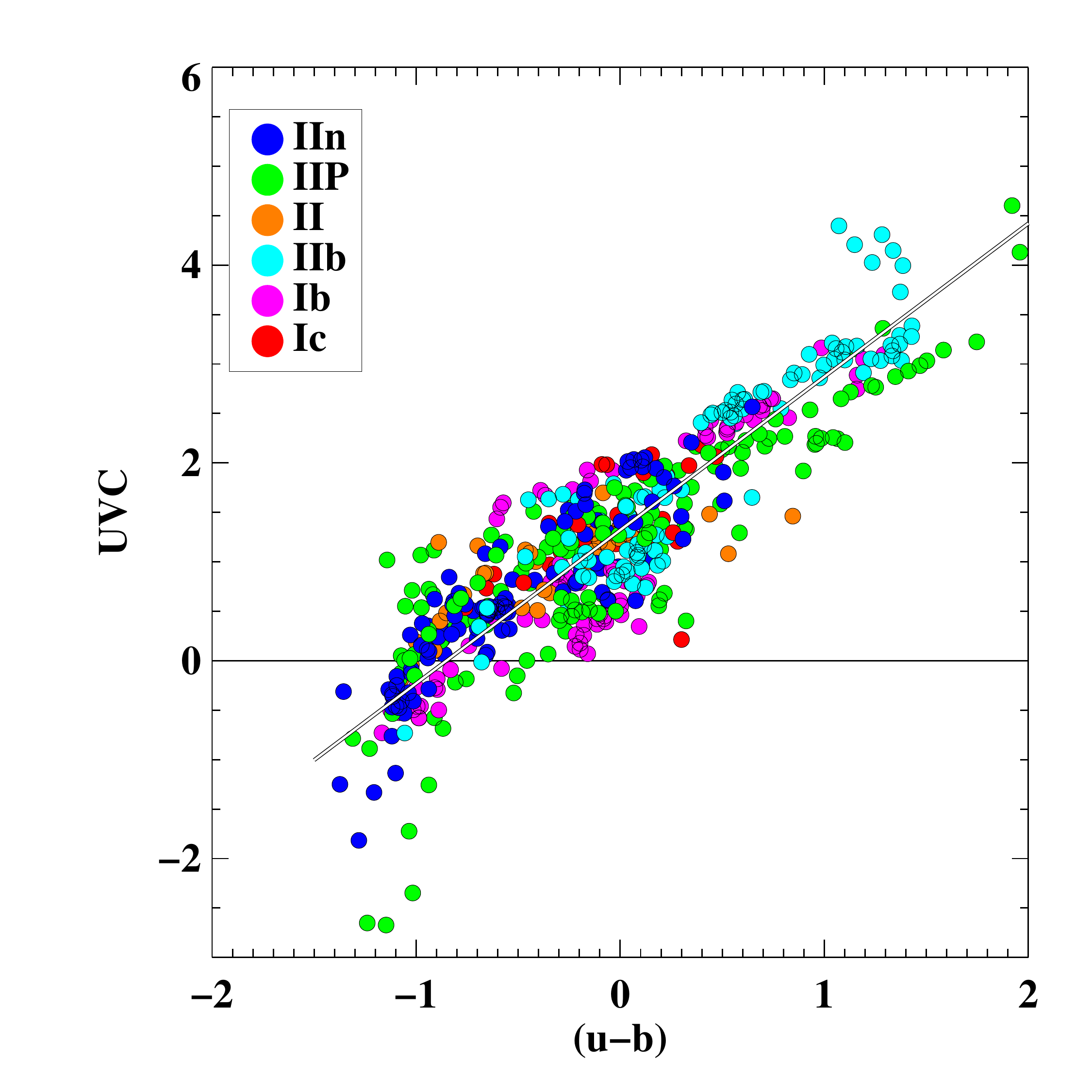}
\includegraphics[width=0.49\textwidth]{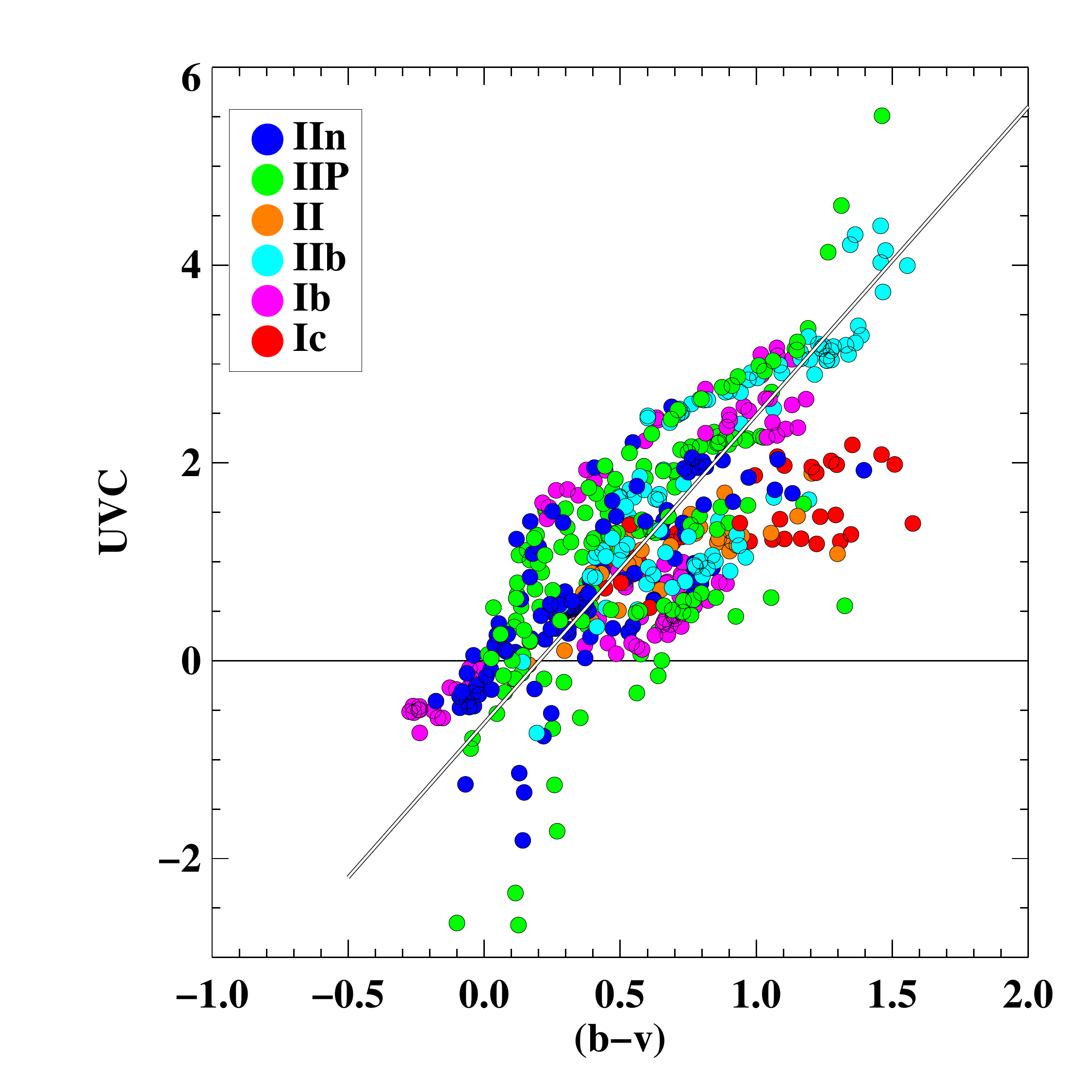}
\caption{\label{colorcorr}  {\em Swift} UV filters observed flux + UV flux correction as a function of optical colors.  The $u$ band here is from {\em Swift}  and has a bluer cut off than ground based filters.  The gray lines represent the Equations \ref{uvcfu} ({\em Left} Panel) \& \ref{uvcfb} ({\em Right} Panel) which are linear fits to the observed data discussed in Section \ref{UV-C}}
\end{center}
\end{figure}

\begin{equation}\label{uvcfu}
UVC = 1.268 +1.529 \times (u-b)\\
\end{equation}

\begin{equation}\label{uvcfb}
UVC = -0.598+3.132\times (b-v)\\
\end{equation}


\subsection{UV Effect on Bolometric Light Curves}\label{UVEff}
As we have demonstrated in Sections \ref{BLC-FC}, \ref{UV-C}, for the CCSNe with a substantial Hydrogen envelope, a substantial amount of a SN's bolometric luminosity lies in the UV regimes at times less $\lesssim 50$ days.  To illustrate how this may effect the bolometric light curve, we present two Type IIP CCSNe models with slightly varying initial parameters in Figure \ref{modelcomp}.  Both models started with a 23 solar mass star evolved until core collapse and then exploded with $5\times10^{50}$ ergs explosion energy \citep{Young06}.  A wind profile was added to each model with a $10^8$ cm/s velocity.  Model A has a dense wind created with a mass loss rate of $10^{-5}$ solar masses per year, and Model B has a mass loss rate of $10^{-6}$ solar masses per year.  Starting immediately after the launch of the shock wave from core collapse, each SN was evolved with the radiation-hydrodynamics code RAGE \citep{Gittings08} and then post-processed with the SPECTRUM code, which uses detailed monochromatic opacities to calculate spectra and light curves \citep{Frey13}.  These models demonstrate that the UV and early time bolometric light curves are very sensitive to the initial progenitor profile and are a valuable addition to constrain models.  At these early times, the optical and light curves are similar but mostly fainter than the UV, and where the UV is dominant we see that small variances in these light curves are reflected by significant changes in the bolometric light curve.  We see this in Figure \ref{modelcomp} ({\em Right}) where the bolometric light curve for model A has a much brighter, narrower peak than model B which evidences a more gradual peak followed by a sharp decline.   This suggests that to both accurately model the bolometric light curve and the underlying progenitor properties we must be able to incorporate this data.

\begin{figure}[h!]
\begin{center}
\includegraphics[width=1\textwidth]{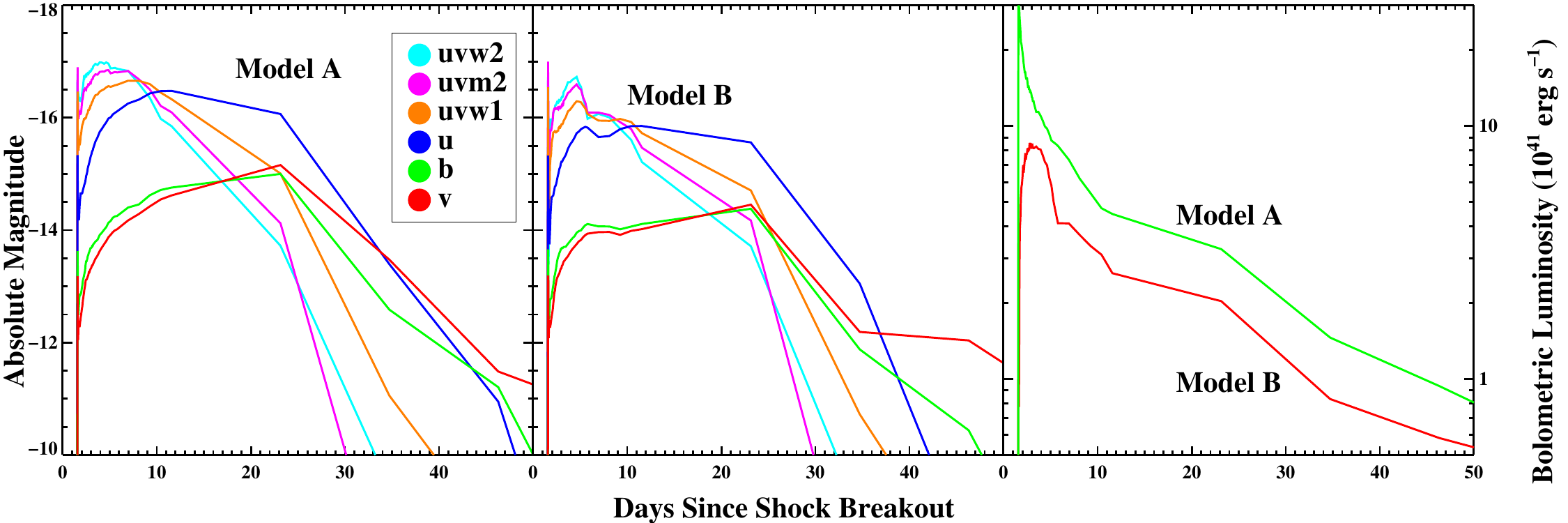}
\caption{\label{modelcomp} A comparison of two model light curves produced from \citep{Frey13}.  These are model runs from a 23 solar mass star, $5*10^{50}$ ergs explosion energy, and mass loss rates of $10^{-5}$/$10^{-6}$ solar masses per year for Model A and B respectively.}
\end{center}
\end{figure}


\begin{figure}[h!]
\begin{center}
\includegraphics[height=1\textheight]{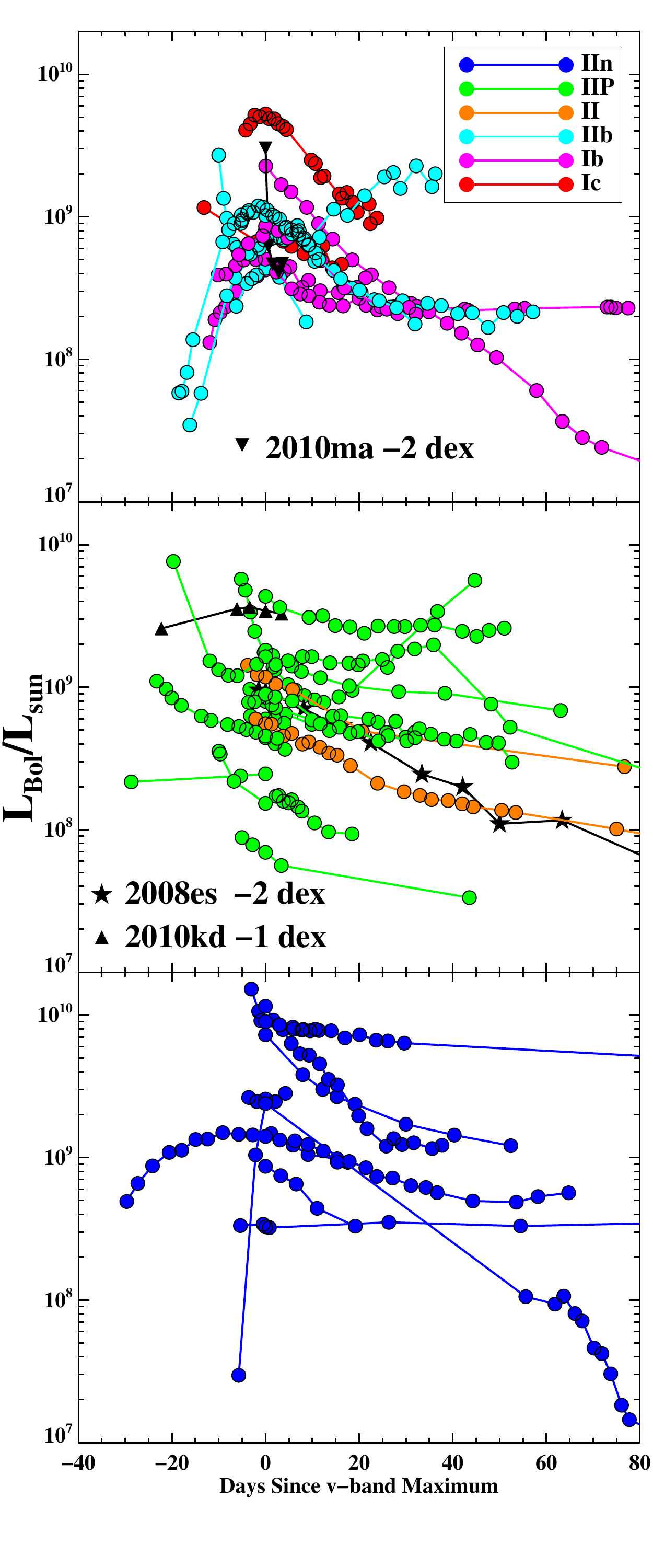}
\caption{ Bolometric light curves  from the {\em Swift} data arranged by SNe subtypes.  The two SLSN 2008es and 2010kd and the GRB-SN 2010ma have had their brightness reduced by a scale factor of 100, 10, and 100 respectively  to bring them in line with the rest of our sample.  \label{blocs}}
\end{center}
\end{figure}


\subsection{Light Curve and Blackbody Behavior}\label{BBBehave}
When we examine the properties of the best-fit blackbodies we see a number of characteristics that are shared across our observed SNe.  For the UV-bright SNe we see that {\em Swift}'s observed peak UV brightness' have temperatures at above $1.5-2 \times 10^4$ Kelvin, and because we often miss the true peak UV brightness (Section \ref{uvrise}) which happens very rapidly, this may serve as a lower bound for maximum temperature in the IIP and IIn SNe where this is the case.  After {\em Swift}'s initial observations, we find that this temperature tends to drop rapidly.  This is due to the cooling of the initial shock breakout.  In the rase case where we catch this tail in the Ib/c/IIb sample (2010jr, 08ax) we see this same behavior if on a shorter timescale due to the lack of a thick Hydrogen envelope.   As our Type IIP's enter the plateau phase we see the best-fit temperatures cluster around a 4500-6000 K values.  As Type IIn's tend to be very UV bright we find that they also tend to fit for higher temperatures and have more variability in their cooling curves, with occasional re-brightening evidenced that is constant with CSM interaction re-heating the ejecta.  Type IIns have been seen to be very UV bright at months and even years after explosion in several cases \citep{Smith09,Stritzinger12} and this general behavior pattern.at early times is consistent with the late time picture of the SN shockwave interacting with a dense progenitor wind or mass-loss/shell ejection events. 

\section{Conclusion}\label{Conc}
The UV properties of CCSNe are diverse, and depend heavily upon subtype.  However, typically the UV's contribution is most important early on, in the $\lesssim100$ days after shock breakout out when the photosphere is still quite hot.  In the rare cases where the UV bands contribute significant flux at late times, it tends to be in IIn SNe where CSM interaction shock heats and excites the gas.   The behavior by subtype does appear to be more homogeneous.  The Type IIn SNe in our sample are the most varied by subtype - while they tend to be UV bright, their behavior varies significantly in other ways such as the duration they are able to be observed and decay rates.  On the other hand the Type IIP as a class is the most homogenous, and is well characterized by a linear decline until $\sim 10-20$ days after v-band max at which point the UV light curves settle into a plateau several magnitudes below the optical \citep[c.f.][]{Bayless13}.  The  IIb/Ib/c SCCSNe fall somewhere in between in terms of homogeneity - they have more individual variation than the IIP, but are more cohesive as a group even considering that we grouped all 3 subtypes together for our purposes.  As a class they are all UV-faint with UV light curves that have a similar shape to the optical but several magnitudes fainter and with a slightly flatter shape.  In several rare cases (2008ax, 2010jr) evidence of a shock breakout cooling tail is evident, and it is in these cases only where we tend to see UV-bright behavior.  These observations raise a number of questions at the moment for which early time observations are crucial, and are now becoming possible given the advent of extremely high cadence SNe surveys now coming online.  \\

When computing bolometric light curves from this sample, we find that {\em Swift}'s observations do a very good job in the first $\sim 50-100$ days in most cases, albeit with a number of caveats.  In the case of Stripped-CCSNe, the SNe's lack of a hydrogen envelope means that the blackbody approximation breaks down much more rapidly than for Hydrogen-rich SNe at the same time as the lack of UV flux makes it harder for {\em Swift} to both measure and fit the SNe light curve.  For other CCSNe at late times, when the UV flux is faint, additional observations red-ward of the UVOT band passes are required in order to better constrain the SED shape.  Nevertheless, at these early times for the IIP and IIn subtypes we find that up to $\sim 75-80 \%$ of the bolometric flux is in the UV at the brightest of observations, and {\em Swift}  can reduce this UV-extrapolation by a factor of 3 or more compared to ground based observations \citep[c.f.][]{Bersten09}.  Using these objects we calculate empirical  Bolometric and UV corrections for use in bolometric light curves calculated from ground based data.  \\ 

We gratefully acknowledge the contributions from members of the {\em Swift} UVOT team at the PennsylvaniaState University (PSU), University College London/Mullard Space Science Laboratory (MSSL), andNASA/Goddard Space Flight Center. This work is sponsored at PSU/ and Southwest Research Institute by the NASA ADP grant  NNX12AE21G.  This research has made use of the NASA/IPAC Extragalactic Database (NED) which is operated by the Jet Propulsion Laboratory, California Institute of Technology, under contract with the National Aeronautics and Space Administration. Report Number LA-UR-13-21329.

\bibliography{apj-jour,myBib}{}
\end{document}